\pdfoutput=1 
\documentclass[conference,compsoc]{IEEEtran}
\usepackage{graphicx}
\usepackage{xcolor}
\usepackage{enumitem}
\usepackage{amsmath}
\usepackage{makecell}
\usepackage{subfig}
\usepackage{multirow}
\usepackage{tabularx,booktabs}
\usepackage{siunitx}
\usepackage[hidelinks]{hyperref}
\usepackage{xurl}
\usepackage{tablefootnote}
\newcolumntype{Y}{>{\centering\arraybackslash}X}
\usepackage{threeparttable}
\usepackage{comment}

\hypersetup{                    
	colorlinks=false,
	linkcolor={black},
	citecolor=black,
	urlcolor={blue!70!black}
}

\newcommand{\ignoreme}[1]{}

\begin{document}

\newcommand{\parhead}[1]{\noindent\textbf{#1}}
\title{AquaSonic: Acoustic Manipulation of Underwater Data Center Operations and Resource Management}

\author{\IEEEauthorblockN{
Jennifer Sheldon\IEEEauthorrefmark{1},
Weidong Zhu\IEEEauthorrefmark{1},
Adnan Abdullah\IEEEauthorrefmark{1},
Sri Hrushikesh Varma Bhupathiraju\IEEEauthorrefmark{1},\\ 
Takeshi Sugawara\IEEEauthorrefmark{4},
Kevin R. B. Butler\IEEEauthorrefmark{1}, 
Md Jahidul Islam\IEEEauthorrefmark{1}, 
Sara Rampazzi\IEEEauthorrefmark{1} 
}
\IEEEauthorblockA{\IEEEauthorrefmark{1}University of Florida;
\IEEEauthorrefmark{4}The University of Electro-Communications
}
}
\maketitle
\begin{abstract}
Underwater data centers (UDCs) hold promise as next-generation data storage due to their energy efficiency and environmental sustainability benefits. While the natural cooling properties of water save power, the isolated aquatic environment and long-range sound propagation characteristics in water create unique vulnerabilities which differ from those of on-land data centers. Our research discovers the unique vulnerabilities of fault-tolerant storage devices, resource allocation software, and distributed file systems to acoustic injection attacks in UDCs. With a realistic testbed approximating UDC server operations, we empirically characterize the capabilities of acoustic injection underwater and find that an attacker can reduce fault-tolerant RAID 5 storage system throughput by 17\% up to 100\%.  
Our closed-water analyses reveal that an attacker can (i) cause unresponsiveness and automatic node removal in a distributed filesystem with only 2.4 minutes of sustained acoustic injection, (ii) induce a distributed database's latency to increase by up to 92.7\% to reduce system reliability, and (iii) induce load-balance managers to redirect up to 74\% of resources to a target server to cause overload or force resource colocation. 
Furthermore, we perform open-water experiments in a lake and find that an attacker can cause controlled throughput degradation at the maximum allowable distance of 6.35 m using a commercial speaker.
We also investigate and discuss the effectiveness of standard defenses against acoustic injection attacks. Finally, we formulate a novel machine learning-based detection system that reaches 0\% False Positive Rate and 98.2\% True Positive Rate trained on our dataset of profiled hard disk drives under 30-second FIO benchmark execution. With this work, we aim to help manufacturers proactively protect UDCs against acoustic injection attacks and ensure the security of subsea computing infrastructures. 
\end{abstract}

\IEEEpeerreviewmaketitle

\section{Introduction}
Data centers play a crucial role in handling and storing vast amounts of data to serve the requirements of different applications owned by private individuals, enterprises, and government institutions. With the increasing interest in environmental sustainability and the surging data center market
due to a recent spike in demand for AI~\cite{datacenterAI} and cloud computing~\cite{datacentercloud}, companies are actively seeking alternative methods to improve energy
efficiency and to reduce operating costs. To this end, Microsoft~\cite{microsoftunderwaterdatacenterarticle}, Subsea Cloud~\cite{subseacloud}, and Offshore Oil Engineering Company~\cite{COOECunderwaterdatacenterarticle}, among others, have already deployed successful prototypes and released in the market underwater data centers (UDCs). Typical UDCs have submerged structures with metal pressurized vessels containing server racks and filled with nitrogen gas to prevent corrosion~\cite{jbod,microsoftunderwaterdatacenterarticle}, which have demonstrated significant advantages due to the natural cooling
properties of water, space efficiency, and renewable energy integration~\cite{ieeeunderwaterdatacenter}.

\begin{figure*}[t]
    \centering
    \includegraphics[width=\textwidth]{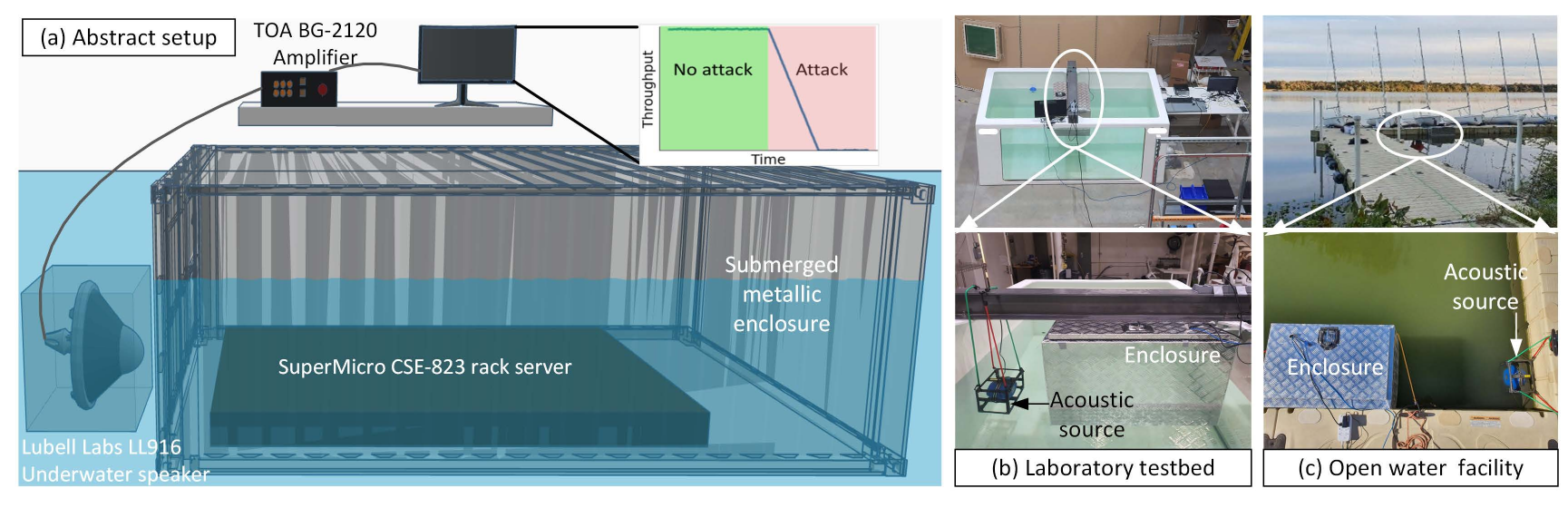}
    \vspace{-2mm}
    \caption{(a) An overview of our experimental setup is shown; a rack server in a RAID 5 configuration is placed inside a $0.9\times0.6\times0.5$ m metal enclosure, while acoustic attacks are carried out using an underwater speaker. (b) The indoor experiments are conducted in a $1.2 \times 3.0 \times 1.5$ m laboratory water tank. (c) The studies are extended to higher volume levels and longer attack distances at an open-water testing facility.}
    \label{fig:overview}
\end{figure*}

Several attack vectors have been studied for in-land data centers, but they generally involve installing malware~\cite{chung2019availability, libri2020paella, abdelsalam2018malware}, target data center networks\cite{liu2010new, anwar2014can,chen2016forwarding}, rely on hardware colocation of virtual machines (VMs),
for eavesdropping~\cite{zhang2012cross, inciseriously, ristenpart2009hey}, or require the attacker to gain physical access and tamper with components inside the data center~\cite{islam2018some, islam2017exploiting, islam2018ohm}. 
More recently, Sheldon~et~al.~\cite{sheldon2023deep} built upon previous works, which explored in-air Denial-of-Service (DoS) attacks on HDDs~\cite{bolton2018blue, shahrad2018acoustic}, by showing that strong sound injection attacks at a resonant frequency of hard disk drives (HDDs) deployed in submerged enclosures can cause throughput reduction and application crashing. While these previous works demonstrated the possibility of DoS attacks on a single disk in air and water environments,
it remains unexplored if such acoustic injection attacks can be used to affect critical data center operations and resource management necessary to ensure the reliability and efficiency of such infrastructure. 
In this paper, we perform modulated acoustic injection attacks in a controlled testbed and real-world open-water scenarios to characterize and quantify an attacker's ability to manipulate complex operations within underwater data centers leveraging the capability of acoustic attacks to influence multiple storage devices simultaneously. Specifically, we test the resilience of fault-tolerant storage techniques such as RAID and investigate attackers' ability to gain fine-grained control over geo-distributed database performance and latency (e.g., CockroachDB~\cite{TSM2020}), distributed filesystem node allocation and replication (e.g., Hadoop Distributed Filesystems (HDFS)~\cite{HADOOP}), and resource allocation (e.g., OpenNebula~\cite{milojivcic2011opennebula}). 
We also show how modulated injection can be used to control the latency of real-world data center workloads such as Microsoft SNIA~\cite{SNIA} at different volumes of injected sound. Our evaluation of full-HDD and Hybrid Solid State Drive (SSD) cache - HDD~\cite{hybridRAID, niu2018hybrid,WLC2020} architectures deployed in current data centers shows that, even if the SSD cache is almost immune to acoustic injection, the attack still increases latency. In fact, for all our tested cache sizes (0.5, 1, 1.5, and 2 GB),  we found that the random-write workload latency rose from the 1 to 200 ms range to the 200 to 800 ms range. 

In our load manipulation analysis, we demonstrate how acoustic injection induces a resource manager (OpenNebula) to assign a minimum of $58\%$ to a maximum of $74\%$ of VMs to a target server during injection. Such assignment manipulation can be used to force benign user applications to be colocated with malicious tenants in compromised servers or to indirectly overload specific servers. In extreme cases, we also cause virtual machines to run into permanent deadlock after the acoustic injection has stopped, which can stop critical processes and corrupt sensitive data. 
Finally, we evaluate our attack in open-water scenario and demonstrate that the attack can achieve $61\%$ throughput drop with the speaker placed $6.35$ m away from the enclosure\footnote{The testing distance was limited to $6.35$ m by the open-water dock used for the metal enclosure anchorage, not by limitations of the setup (see ~\autoref{fig:overview}). More information can be found at https://cpseclab.github.io/aquasonic/}. 

In light of our characterization of the attacker's capabilities, we discuss how simple defenses, such as the use of sound absorbing materials, should be carefully considered to avoid dangerous heating which can impact the performance of the data center servers in submerged enclosures. To address this, we develop and evaluate a novel proof-of-concept machine learning (ML) based defense that
detects whether acoustic injection has occurred based on throughput analysis along with information on acoustic vibration patterns. 

\vspace{1mm}
\noindent

\noindent
In summary, this paper includes the following contributions:

\begin{itemize}[leftmargin=1.3em,topsep=1pt,noitemsep]

\item We characterize submerged data center storage devices' vulnerability to acoustic injection. We perform real-world testing in a simplified testbed and open-water scenarios, using a server enclosed in a submerged metal structure. Our evaluation shows remote throughput manipulation more than 6 meters away from the enclosure. We also build a preliminary simulation model for evaluating different subsea UDC structures.

  \item We demonstrate how an adversarial attacker can affect the performance and reliability of distributed databases and filesystems in UDCs by performing acoustic injection on CockroachDB\cite{TSM2020} and a server running the DFSIO~\cite{DFSIO} file access benchmark on HDFS~\cite{HADOOP}. We are able to increase CockroachDB's latency up to $92.7\%$ and block Hadoop from accessing data within 144 seconds of injection.
  \item Furthermore, we demonstrate how attackers can manipulate data center resource allocators, such as OpenNebula, and force up to $74\%$ resource reassignment while circumventing VM placement policy-based defenses.
  \item We investigate the effectiveness of standard defenses against acoustic attacks, such as using sound-absorbing material, SSD-hybrid architectures, active noise cancellation, feedback controllers, and sensor fusion. 
  \item We propose and evaluate a novel ML-based defense that models patterns of storage device throughput and identifies attacks based on anomaly estimation across multiple spatially close storage devices.
  Comprehensive evaluations in our real-world testbed conditions leveraging the FIO sequential write benchmark show that the proposed defense achieves $0\%$ False Positive Rate and $98.2\%$ True Positive Rate in detecting simultaneous degradation caused by the attack.
  \end{itemize}
\vspace{1mm}
Overall, our analysis begins to unveil hardware and software vulnerabilities that are unique to submerged environments, while revealing new design flaws in traditional data centers fault tolerance storage devices and resource management systems. This work aims to help manufacturers and designers of UDC infrastructures promptly address those security risks before they become widespread.

\section{Background}
\label{sec:background}
\subsection{Data Center Architectures}
\label{sec:background:datacenter}
Users, companies, and organizations are shifting their data and business to the cloud with unprecedented speed, especially since the COVID-19 pandemic which mandated remote work and study~\cite{MS775}. This surge fueled the ongoing expansion of data center services, which continues to this day. For example, at the time of writing, data center construction is projected to reach \$49 billion by 2030 with a power consumption estimated at 35 gigawatts only in the United States~\cite{datamarket}.
To satisfy the high demand for data storage and computing resources, data centers comprise high-speed processors, servers, network switches and routers, and large-scale storage systems~\cite{DCCOMPO}.

In addition, data center architectures deploy fault tolerance techniques, resource allocation, load and workload balancing processes, and storage systems management tools to ensure the reliability, performance, and efficiency of the infrastructure when handling data. In this work, we focus on understanding and evaluating how such supporting resources are manipulated by acoustic attacks affecting storage devices in the context of underwater data center deployment.

\vspace{1mm}
\noindent
\textbf{Storage Devices in Data Center.}
Data centers providers, such as Alibaba Pangu~\cite{PANGU}, Microsoft Azure~\cite{CWO2011}, and Meta~\cite{MLR2014}, use multiple types of storage devices~\cite{HDDSSDDC}, including HDDs and SSDs, to satisfy their demands on storage capacity and performance.
Although SSDs have significant advantages in performance and reliability~\cite{SML2017}, HDDs remain  
the main components of current data center storage systems due to the following reasons.
First, SSDs suffer of limited lifetime due to finite program/erase (P/E) cycles and high \textit{infant mortality} rate~\cite{MKC2012}.
Second, SSDs rely on garbage collection (GC) to deal with their underlying NAND flash memory~\cite{IS2011,YLH2017}
that prohibits in-place updates, greatly degrading performance~\cite{YLH2017,WZW2018} and introducing unpredictability~\cite{HTL2020,LPS2021}.
Finally, SSDs are still expensive~\cite{WLC2020,LZL2019} compared to their equivalent HDDs.
For example, one terabyte of HDD costs around 35 dollars, whereas SSD might run 60 to 80 dollars~\cite{HDDVSSSD}. 
Therefore, HDDs and SSDs co-exist in modern data centers, and SSDs are used mainly as a cache~\cite{WLC2020,CWO2011,MLR2014} of HDDs to boost the performance of the storage system.
In this work, we thus focus primarily on full-HDD and Hybrid SSD cache architectures to evaluate the impact of acoustic attacks on UDCs. 

\vspace{1mm}
\noindent
\textbf{Redundant Array of Independent Disks (RAID).}
Typically data centers deploy storage devices with RAID technology~\cite{CLG1994}
to enable data processing in multiple hard disks simultaneously while providing fault tolerance.
There are six types of configurations in a RAID system.
RAID 0 evenly divides storage space and distributes incoming data in chunks, enabling speed up by simultaneous disk access.
However, RAID 0 offers no fault tolerance, and the failure of one storage device will lead to data loss. 
Therefore, RAID 2 to RAID 6 provides fault tolerance by adding redundancy and using erasure coding techniques.
Specifically, RAID 1 relies on mirroring data -- adding replicas -- across multiple storage devices, whereas RAID 2 to RAID 6 use erasure code to provide fault tolerance without significant storage overhead.
While RAID 2 and RAID 3 are rarely used in practical applications because they chunk data in byte granularity, making processing I/O requests in parallel difficult and incurring poor performance, RAID 5 and 6 are commonly used in enterprise storage systems.
In this work, we consider RAID 5 as the implementation in our experimental setting because widely used in the storage system of data centers~\cite{SZM2023,MJW2012,IS2011}.

\vspace{1mm}
\noindent
\textbf{Resource Allocation Techniques.}
Data centers allocate computing and storage resources on demand.
Unlike provisioning resources on a single machine, a data center allows a flexible and scalable combination of system resources, such as memory, CPU cores, and storage devices.
For example, Infrastructure as a Service (IaaS)~\cite{IAAS} allows users to purchase a specific amount of computing power and storage space to fulfill their demands; storage space in the RAID device is assigned upon request and disaggregated on release~\cite{SZM2023}.
In addition, to achieve high-performance utilization of system resources, data centers optimize the use of their resources by balancing the workload between servers~\cite{loadBalancing}. 
Workloads are distributed to servers based on varying optimization parameters. 
For example, OpenNebula~\cite{milojivcic2011opennebula} assigns virtual machines to hosts based on varying optimization parameters such as number of hosts used, available server resources, or custom algorithms~\cite{OpenNebulaDocs}. 
Through our evaluation of OpenNebula in \autoref{sec:loadManipulation}, we demonstrate how attackers can manipulate resource allocation in data centers to overload servers or force colocation with malicious virtual machines to perform potential attacks such as eavesdropping~\cite{zhang2012cross, inciseriously, ristenpart2009hey}.

\subsection{Acoustic Injection Attacks}
\label{sec:background:acousticInjection}
Acoustic injection attacks involve the use of sound to manipulate the behavior of a target system.
These attacks usually leverage the \textit{resonant frequency}, meaning the natural frequency at which a solid structure oscillates. Several parameters can generate resonance frequencies, including the
elasticity or stiffness of the material from which a system or component is made, the physical dimensions, and the object’s mass distribution. Additionally, complex structures may exhibit multiple resonance
frequencies corresponding to different vibration modes. Thus acoustic attacks consist of transmitting acoustic waves at a frequency that matches the resonant frequency, to efficiently convert acoustic waves into physical vibrations of the target system~\cite{halliday2013fundamentals}.
Researchers have explored acoustic injection attacks against a variety of sensors, including (i) cameras~\cite{ji2021poltergeist, cheng2023adversarial} to induce obstacle misdetection in autonomous vehicles, (ii) accelerometers~\cite{trippel2017walnut}, gyroscopes~\cite{son2015rocking, khazaaleh2019vulnerability} and inertial sensors~\cite{tu2018injected, gao2023exploring} to alter drone locations, and (iii) microphones to stealthily deliver voice commands~\cite{zhang2017dolphinattack}. In particular, Blue Note~\cite{bolton2018blue} used acoustic injection at the resonant frequency of an HDD to vibrate its actuator arm outside the disk track's limits, causing DoS in laptops and security cameras.

More recently, Sheldon et al.~\cite{sheldon2023deep} in a position work showed a DoS attack against a single HDD located in a closed submerged container made of different materials (plastic and aluminum). The attack is successful in short distances (up to 25 cm from the container), causing operating system crashes and I/O timeout errors.
Although this preliminary work pioneered how sound waves in water can induce mechanical vibrations that propagate through solid structures (e.g., the target hard disk drive and container), it was limited in reproducing Blue Note in underwater settings. 

Unlike these previous works, our analysis focuses on leveraging the vulnerability of storage devices to acoustic injection to subtly manipulate complex operations and processes (e.g., resource allocation, fault tolerance techniques) vital for maintaining the reliability of data centers.

\subsection{Acoustic Signal Propagation in Water}
\label{sec:background:waterprops}
Sound waves propagate in water at a much higher velocity over long distances than in air. This is due to the higher density of water, which allows for efficient transmission with minimal loss of energy. In general, the speed of sound is around 1,480 m/s, which is about four times faster than in the air ~\cite{lurton2002introduction}. 
More specifically, sound propagation in water depends on many factors, such as temperature, salinity, depth, seabed relief, currents, and surrounding pressure~\cite{kozaczka2017theoretical}. In shallow water, sound waves reflect from the surface and the floor of the water body, and scatter from suspending particles and bubbles, resulting in multi-path propagation. In deep water instead, pressure and temperature vary with depth which creates multiple layers in the medium and sound bends while propagating through such layers~\cite{kim2012underwater,erbe2022introduction}. High-frequency energy is scattered and absorbed more rapidly by a water medium that allows low-frequency sound to travel longer distances. Additionally, for a given sound frequency, sea water shows typically a higher absorption coefficient than fresh water due to its higher concentration of dissolved minerals~\cite{hovem2007underwater}.
All the experiments in this work are conducted in fresh water scenarios where the acoustic source is kept at a certain depth from the surface and the ambient temperature is maintained.

\begin{figure}[t]
	\centering
	\subfloat[Laboratory testbed]{%
			\includegraphics[width=0.47\columnwidth]{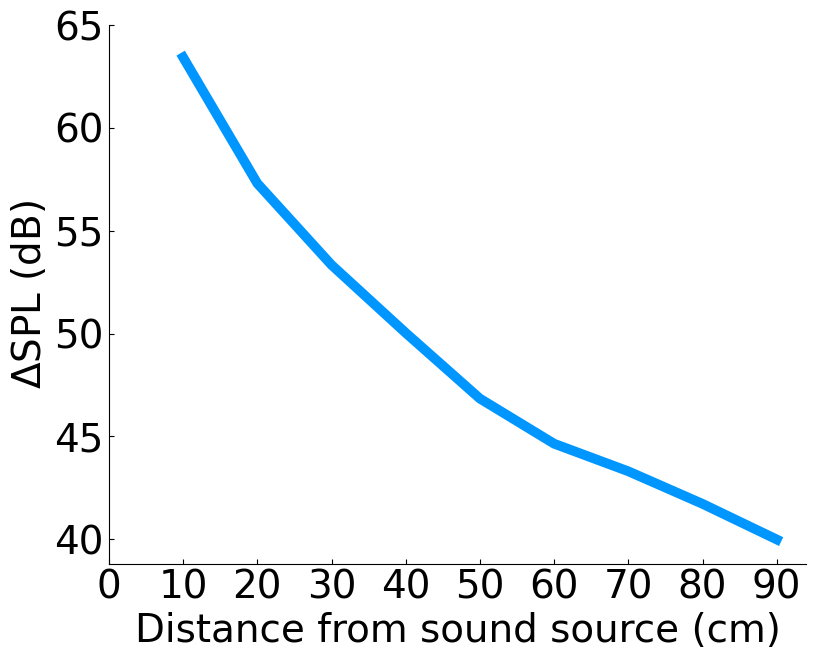}
		}
	\hspace{\fill}
	\subfloat[Open water]{%
		\includegraphics[width=0.47\columnwidth]{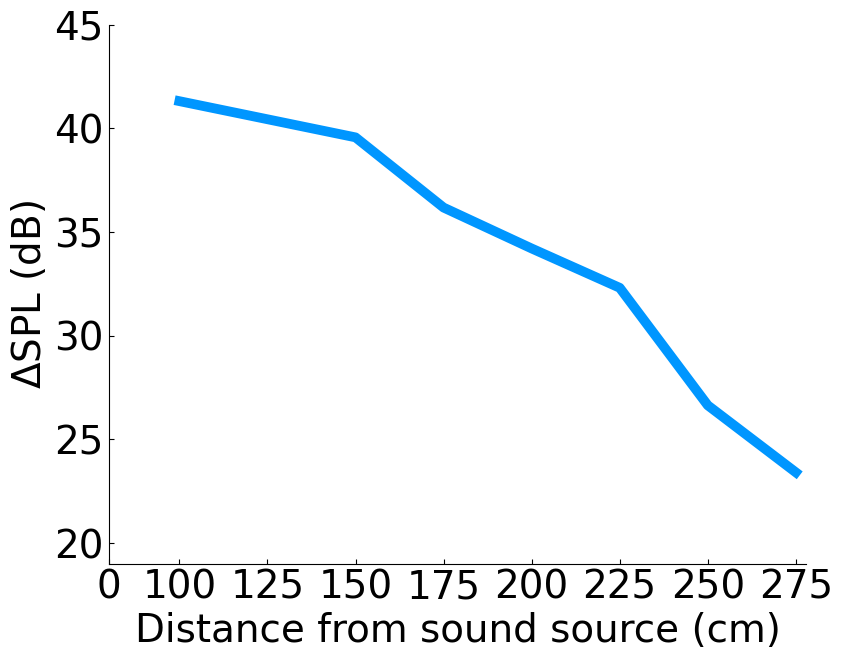}
	}
 \caption{Sound pressure attenuation at increasing distances are shown for closed-water laboratory testbed and open-water scenarios; we find similar trends in both scenarios.}%
\label{fig:splcurve}
\end{figure}

The sound pressure level (SPL) is the typical measure of the intensity or loudness of a sound wave and is expressed in decibels (dB) as the product of medium density, wave velocity, and particle velocity in a specific medium~\cite{pierce2019acoustics}. 
Sound pressure levels in water are about 60 dB higher than the equivalent SPL in air for the same sound source. This means that a sound source producing 100 dB SPL in air, is approximately equivalent to 160 dB SPL sound in water.
In all the experiments in this work, we determine the sound pressure by using a hydrophone and empirically measure
the sound pressure generated by the sound source (e.g., our underwater speaker) at increasing distances (see~\autoref{fig:splcurve}).

\vspace{1mm}
\noindent
\textbf{Mechanical Vibrations Induced by Sound.}
Sound waves can propagate through a fluid to induce mechanical vibrations in a solid. In our experiments, sound waves propagate from the sound source through the water to produce mechanical vibrations in the submerged structure described in~\autoref{sec:expsetup}. Sound propagating through a solid produces various types of waves, including longitudinal and shear~\cite{mason2013physical}. As a simplified example of acoustic propagation in sound, in an infinite isotropic solid (i.e., a solid that is invariant to orientation~\cite{isotropic}), the longitudinal wave velocity is 
\begin{equation*}
    v_{l} = \sqrt{(\lambda + 2 \mu_1)/\rho},
\end{equation*}
while shear wave velocity is
\begin{equation*}
    v_{s} = \sqrt{\mu_2/\rho}.
\end{equation*}
Here, $\lambda$ and $\mu_{1}$ are Lam\'e coefficients, $\rho$ is the density of the material, and $\mu_{2}$ is the shearing modulus, all of which are properties of the solid through which the wave propagates~\cite{lameCoefficients}. By performing acoustic injection at a resonant frequency of a structure, which is also determined by its material properties and shape~\cite{resonantFreq}, we maximize the amplitude of the propagating waves.

\section{Attack Overview and Threat Model}
\subsection{Threat Model}
In this work, we consider an adversary who aims to achieve high-level control over underwater data center infrastructure operations by exploiting the vulnerability of the storage systems to acoustic injection. Unlike the previous works~\cite{bolton2018blue, sheldon2023deep, shahrad2018acoustic} focused on DoS attacks, this work aims to characterize the extent and capability of an attacker to induce at far distances sophisticated and subtle manipulations of data center operations such as redirecting workloads, altering resource allocation, and achieve fine-grade control over latency and throughput of critical distributed applications. We reveal how these manipulations can be pursued without evident and abrupt changes in the process executions, causing storage devices to be automatically removed, forcing automatic node redirection, and dangerous colocation of resources which can overload specific servers.
\vspace{1mm}
\noindent
\textbf{Attacker Knowledge and Assumptions.}
We assume that the adversary has no direct access to the underwater infrastructure and cannot tamper with any hardware, software, or network communication. Meanwhile, the adversary can observe the outcome of the acoustic attack by monitoring benign application instances running on the infrastructure. A less sophisticated attacker without direct access to application instances can also analyze effective signal injection frequencies and volumes by performing small-scale evaluations on enterprise HDDs as we show in this work, or simulating the target's structure using physics modeling simulators. The attacker can also perform a simpler attack by emitting sound waves and continuously sweeping a range of frequencies.
We assume that the attacker owns an underwater speaker capable of generating and controlling the volume and frequency of the acoustic waves.
The attacker is capable of aiming the speaker at the data center location to perform the attack for the desired time span. This can be achieved by mounting the speaker on a rigid structure connected to a boat or using more sophisticated settings, such as a remotely controlled underwater robot\cite{gu2020highly}.
The attacker can also make use of directional speakers such as speaker arrays or Long Range Acoustic Device (LRAD) devices~\cite{LRAD} to transmit focused beams of sounds with a confined range to target specific structure parts.

In order to perform the attack, the adversary should identify the susceptible frequency ranges. This is possible, by observing delays in benign application write requests caused by brief sound injections and different frequencies or by studying similar storage devices and
their resonance frequencies as shown in previous
work~\cite{bolton2018blue, sheldon2023deep}.

\vspace{1mm}
\noindent
\textbf{Attack Scenario.}
\label{sec:attackScenario}
Adversaries can launch the attack several meters away from the underwater infrastructure, depending on their equipment and the susceptibility of the victim system. We first characterize the vulnerability in a controlled scenario using a laboratory testbed in Section~\ref{sec:capability}. In the open-water scenario described in Section~\ref{sec:evaluation:openwater}, we achieve successful attacks at 6.35 meters away using a commercial speaker connected to an amplifier. A well-funded adversary with powerful speakers (e.g., military-grade equipment) can potentially reach further distances as we explored in our simulation scenario in Section~\ref{sec:evaluation:simulation}.

\subsection{Attack Overview}
\label{sec:attackScenarioTestbed}
As shown in ~\autoref{fig:overview}, the attacker uses an underwater speaker to generate modulated sound waves in the form:
\begin{equation} 
s(t) = A \cdot \cos(\omega \cdot t)
\label{eq:sineWaves}
\end{equation}
where A is the amplitude of the sound wave corresponding to the volume, $\omega$ is the angular frequency of the transmitted sound, and t is time. We also define the amplitude of the signal as a decibel sound pressure level (SPL) above the noise level using the formula: $\Delta SPL = SPL_{m} - SPL_{n}$. Here, $SPL_{m}$ is the SPL (in dB) measured in the testing environment with a hydrophone, and $SPL_{n}$ is the environmental noise SPL measured when no sound is being emitted. A higher $\Delta$SPL is associated with a louder volume.

As described in~\autoref{sec:background:acousticInjection},
transmitting acoustic waves at resonant frequencies with sufficient volume can cause mechanical vibrations in the internal components of HDDs (e.g. read/write head and platter), preventing reading and writing operations and consequently causing application crashes. We apply amplitude modulation to induce controlled changes in the behavior of a victim system composed of multiple storage devices in full-HDD and hybrid SDD-HDD architectures as described in Section~\ref{sec:background:datacenter}.

\vspace{1mm}
\noindent
\textbf{Indoor Testbed Specifications.}
To approximate a simplified underwater infrastructure and perform our vulnerability characterization, we use the indoor testbed depicted in ~\autoref{fig:overview}. This testbed consists of a $1.2\times3.0\times1.5$~m water tank filled with fresh water.
An aluminum metal enclosure of $0.9\times0.6\times0.5$~m is used to emulate a real-world data center vessel while a rack server (or rack-mounted server) is used as the target system, as UDCs infrastructures are composed of server racks~\cite{jbod,microsoftunderwaterdatacenterarticle}.

\subsection{Theoretical Analysis}
\label{sec:theoretical}

The sound-induced vibrations in submerged enclosures depend on three main physical phenomena: the propagation of sound through fluids, the force applied by fluids on solid boundaries, and the propagation of mechanical vibrations through solids.

In our attack model, sound waves travel through a body of water to reach the enclosure and then propagate to the victim server's solid structure containing the hard disk drives. Sound propagation in fluids is generally modelled using the equations~\cite{jensen2011computational}:

\begin{equation}
\begin{cases}
    \frac{1}{\rho}\nabla^{2}p_{t}(x) - \frac{k^{2}}{\rho}p_t(x) = 0 \\
    k = \frac{\omega}{c} \\
\end{cases}
\end{equation}

Where $\rho$ is the density of the analyzed fluid, $p_{t}(x)$ is the total pressure in the fluid at a location \textit{x} assuming the acoustic source at position 0, $\omega$ is the angular frequency of the source sound, and c is the speed of sound, which depends on temperature, salinity, and depth of the fluid.

Sound waves attenuate with distance from the source, and this attenuation can be modeled as: 
\begin{equation}
   A = A_{0}e^{-{\alpha}x}
   \label{eq:attenuation}
\end{equation}
where \textit{A} typically is represented by the root mean square amplitude (RMS) of the wave at distance x, $A_{0}$ is the RMS when x is zero, and $\alpha$ is the attenuation coefficient of the fluid, which depends on the sound frequency.

The force applied by a sound propagating through a fluid and encountering a solid can be approximated using the equations~\cite{harari1996recent}:

\begin{equation}
    \begin{cases}
        \frac{-\boldsymbol{n}}{\rho}{\nabla}p_{t} = -\boldsymbol{n} \cdot (\boldsymbol{u_{tt}})\\
        F_A = p_t\boldsymbol{n}
        \label{eq:solidLiquidBoundary}
    \end{cases}
\end{equation}

Where $\rho$ is the density of the fluid, $\boldsymbol{n}$ is the surface normal of the solid, $\boldsymbol{u_{tt}}$ is the acceleration vector of the solid, $p_{t}$ is the total acoustic pressure, and $F_A$ is the fluid load on the solid boundary which depends on the distance of the sound source and its attenuation.

The force applied by the sound pressure induces vibrations in the submerged metal enclosure, which can be expressed as~\cite{howe1998acoustics}:

\begin{equation}
\
    \rho_{s}\frac{\delta^{2}\boldsymbol{u}}{{\delta}t^2} = \boldsymbol{F_{v}}{\nabla_{X}} {\cdot} P
\end{equation}

Where $\rho_s$ is the density of the solid, $\boldsymbol{u}$ is the displacement vector of each point in the solid material, $\boldsymbol{F_{v}}$ is the vector force per unit volume applied to the solid. Here, the force is calculated upon $F_A$ from Eq.~\ref{eq:solidLiquidBoundary} at the fluid-solid boundary. 
Finally, the propagation of mechanical vibrations at the interface between solids of different densities, such as the vessel structure, the internal server racks, and hard disk configurations, depends on complex interactions based on the specific materials of each component, boundary conditions, and other factors such as reflection, transmission, and mode conversion~\cite{feng2021strain}.  
When mechanical vibrations caused by the injection encounter an interface between two different materials, part of the wave is reflected back into the original material, while part of it is transmitted into the second material~\cite{ingard1959influence}. The coefficients of reflection and transmission can be calculated using the acoustic impedances of the two materials. This can be represented by the following simplified equations:

\begin{equation}
    \begin{cases}
        R = \frac{{Z_2 - Z_1}}{{Z_2 + Z_1}} \\
        T = 1 + R = \frac{{2Z_2}}{{Z_2 + Z_1}}
        \label{eq:solidSolidBoundary}
    \end{cases}
\end{equation}

Where (R) and (T) are the reflection and transmission coefficients, respectively, and ($Z_1$) and ($Z_2$) are the acoustic impedances of the first and second materials. In addition to reflection and transmission, mode conversion can occur at the interface. For instance, an incident longitudinal wave can generate reflected longitudinal and transverse waves, and transmitted longitudinal and transverse waves. The same applies to an incident transverse wave based on the angle of incidence of the wave, the acoustic impedances, and the frequency of the wave.  At the resonance frequency ranges, meaning where the frequency of the sound wave matches the natural oscillating frequency of the solids in contact with each other, the mechanical impedance will be lower, meaning less force will be needed to propagate the wave at the target object (e.g., the storage system) and cause vibrations at a given velocity and intensity directly proportional to the sound pressure level.

\subsection{Sound-induced Vibrations at the Disk}
\label{sec:PES}
\begin{figure}[t]
\includegraphics[width=\columnwidth]{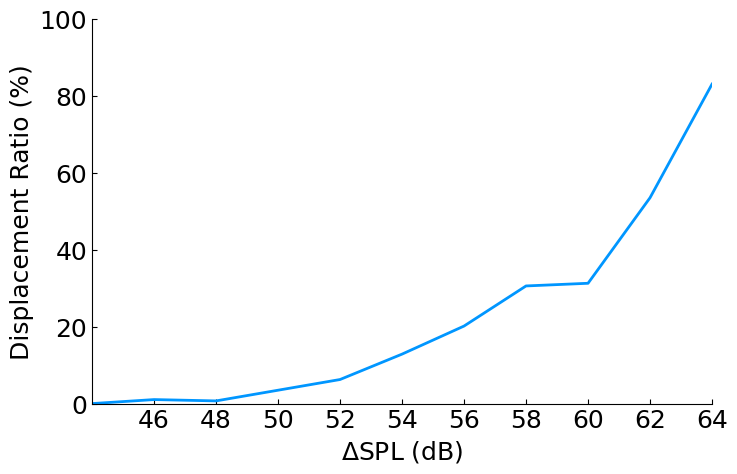}%
\vspace{-1mm}
\caption{PES displacement ratio at different $\Delta SPLs$}
\label{fig:PES}
\end{figure}

Based on our theoretical analysis, we verify the vibration propagation in a storage device located in our server enclosed in the submerged metal structure in our indoor testbed. To measure the resulting vibrations caused by the sound, we extract the Position Error Signal (PES) as described in previous work~\cite{kwong2019harddriveofhearing}. The PES measures the deviation of the read/write head from the center of the track, thus we placed a 500 GB Seagate Barracuda HDD housed in a SuperMicro CSE-823 rack server in our indoor testbed at 6 cm from the edge of the submerged enclosure. To evaluate the deviation of the read/write head, we inject a tone at the HDD resonance frequency (5.1 kHz) at increasing volumes (46 to 64 dB $\Delta SPL$). We leverage the Servo Batch Test~\cite{SERVOCMD} in the Seagate terminal command set to get the PES data from the HDD where each test has 296 revolutions. \autoref{fig:PES} shows that the average displacement ratio increases from $0\%$ to $83\%$ at increasing volumes. This verifies that sound tones at the resonance frequency induce vibrations in the read-write head and platter of the disks by vibration propagation, which is proportional to the acoustic pressure (intensity of the sound, volume) generated by the injection. In this work, we show how an attacker can control the degree of induced effect in applications by varying the injected sound volume.

\section{Vulnerability Characterization}
\label{sec:capability}
In this section, we determine whether an attacker can exploit the resonant frequency of a server composed of multiple storage devices in RAID 5 configuration in a submerged metal enclosure. 
We also evaluate the attacker's ability to maintain fine-grained control over throughput and latency to perform subtle attacks. Through this evaluation, we then quantify the attacker's limitations and capabilities in attacking a data center high-level operations using acoustic injection in ~\autoref{sec:evaluation}.
\\\parhead{Experimental Setup.}
\label{sec:expsetup}
We perform our characterization analysis using our indoor testbed described in Section~\ref{sec:attackScenarioTestbed}.
The sound source is a Lubell Labs LL916 speaker~\cite{lubellspeaker} used for commercial applications (e.g., delivering verbal instructions to divers and swimmers).
As target system located in the submerged metal enclosure, we deploy a SuperMicro CSE-823 rack server~\cite{SuperMicroChassis} running Ubuntu 22.04 with 4 Seagate Exos 7E2 1TB SATA enterprise HDDs~\cite{HDD} used in datacenters in a RAID 5 full-HDD configuration and an Intel D3-S4510 Series application SSD~\cite{SSD}. Hybrid storage architectures are explored in Section~\ref{sec:hybrid}.

\subsection{Resonant Frequency Identification}
\label{sec:capability:frequency}
As the first step in our attack characterization, we determine whether attackers can find and exploit the resonant frequencies of our target server by observing the response to a simple sound wave at constant volume.
 To accomplish this goal, we place the running server in the submerged metal enclosure such that the front of the server is 3 cm away from the enclosure. We then inject sound waves at 150 dB SPL (equal to 34 dB $\Delta$SPL) with frequencies ranging from 100 Hz to 12 kHz with 100 Hz increments every 5 seconds. While playing these tones, we run FIO, an I/O tester tool~\cite{FIO}, with sequential read and write throughput benchmarks to determine which frequencies, if any, can cause measurable throughput degradation~\cite{sheldon2023deep}. We run both the read and write sweeps three times and classify an average performance decrease of more than 20\%.
 
\begin{figure}[t]
\includegraphics[width=\columnwidth]{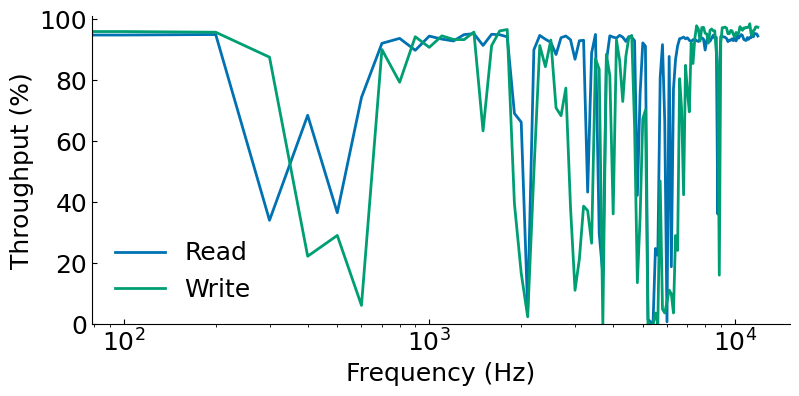}%
\vspace{-1mm}
\caption[]{RAID 5 throughput as a percentage of the baseline average at varying frequencies based on a frequency sweep from 100 Hz to 12 kHz at 100 Hz intervals.{\footnotemark}}
\label{fig:freqSweep}
\vspace{-3mm}
\end{figure}

As shown in ~\autoref{fig:freqSweep}, the measured RAID 5 throughput drops at varying frequencies. Per our theoretical analysis, this is likely because the components inside the four HDDs in the RAID configuration and server structure have varying resonant frequencies  (around 2.0, 3.7, 5.1--5.3, and 8.9 kHz, see \autoref{fig:freqSweep}) with higher frequencies hitting harmonics of the resonant frequencies. The consistent throughput degradation between 5.1--5.3 kHz is 
a good attack target, and we use 5.1 kHz for the rest of our experiments in this work.

\footnotetext{Updated data.}
\subsection{Controlled Injection}
\label{sec:volumeChar}
\begin{figure}[t]
	\centering
	\subfloat[Laboratory testbed]{%
		
			\includegraphics[width=0.47\columnwidth]{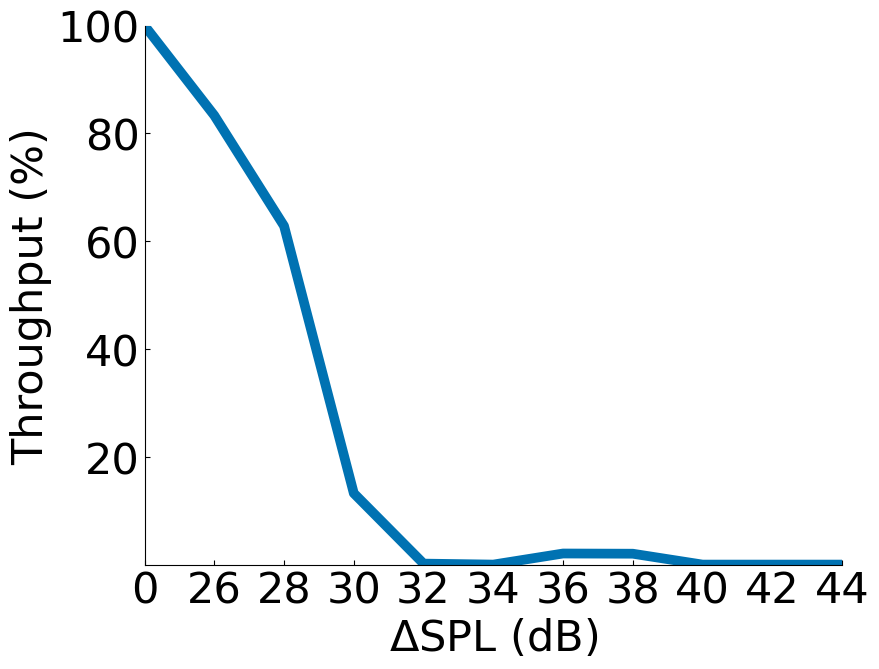}
		}
	\hspace{\fill}
	\subfloat[Open water]{%
		\includegraphics[width=0.47\columnwidth]{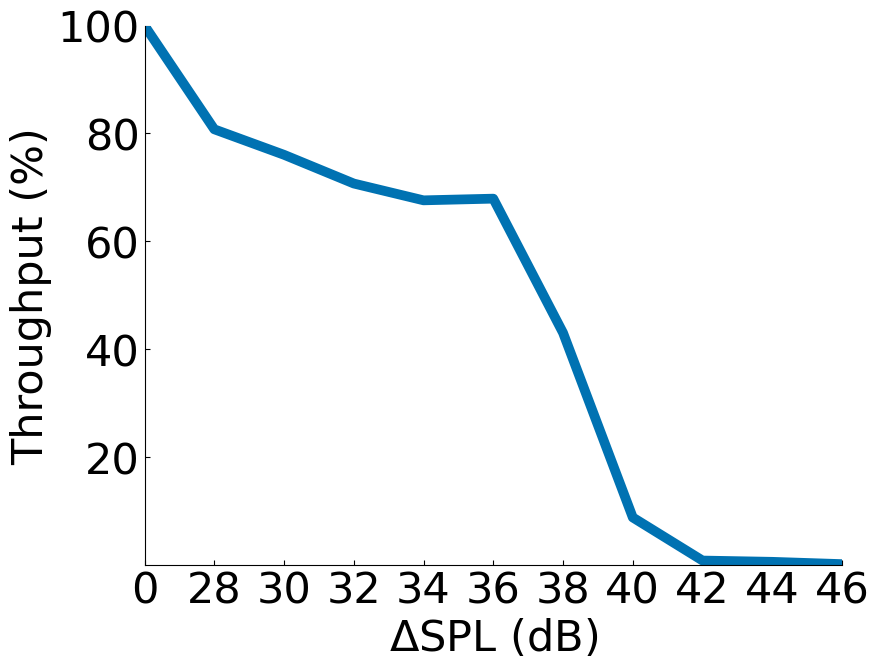}
	}
 \caption{RAID 5 write throughput at increasing sound pressure from the baseline environmental noise (116 dB SPL for the testbed, and 114 dB SPL for the open water scenario) at $5.1$\,kHz injection frequency. The distance between sound source-enclosure was set to $6$\,cm for our laboratory testbed and $30$\,cm for open water scenario.}
\label{fig:volumes}
\vspace{-3mm}
\end{figure}

To demonstrate how an adversary can control the severity of throughput degradation to perform a more subtle attack, we place the sound source speaker at a fixed distance of 6 cm from the outer surface of the submerged enclosure containing the target server. We then play sounds at the resonant frequency of 5.1 kHz identified in~\autoref{sec:capability:frequency} with 2 dB SPL increment starting from 26 dB $\Delta$SPL. During these acoustic injections, we record the average RAID 5 throughput of three 30-second runs of the FIO sequential write benchmark. As shown in~\autoref{fig:volumes}, the attacker can drop the throughput between 17 and 100\% by varying the acoustic injection volume between 26 and 32 dB above the noise level (116 dB SPL in our indoor testbed scenario). 

Thus, the attacker can subtly control the throughput of storage devices in the target server by varying the injection volume \textit{A} (See Eq.~\ref{eq:sineWaves}). By associating the volume with sound source-to-target distance using our empirical results to approximate the attenuation constant of water as shown in ~\autoref{fig:splcurve}, we also determine that the attacker can control the RAID 5 throughput by varying the distance from which they inject a constant volume signal.

We validate our analysis as well in our open-water scenario. As depicted in~\autoref{fig:volumes}, the open water scenario required a higher injection volume to reach similar throughput degradation. This is likely because, due to the absence of a fixed anchorage in our open water setup, we weighted the metal enclosure using bags of sand to reach the submerged level required. We believe this procedure may have influenced the resulting vibrations in the enclosure.

\subsection{Acoustic Injection Points}
\label{sec:injectionPoints}
\begin{figure}[t]
\includegraphics[width=\columnwidth]{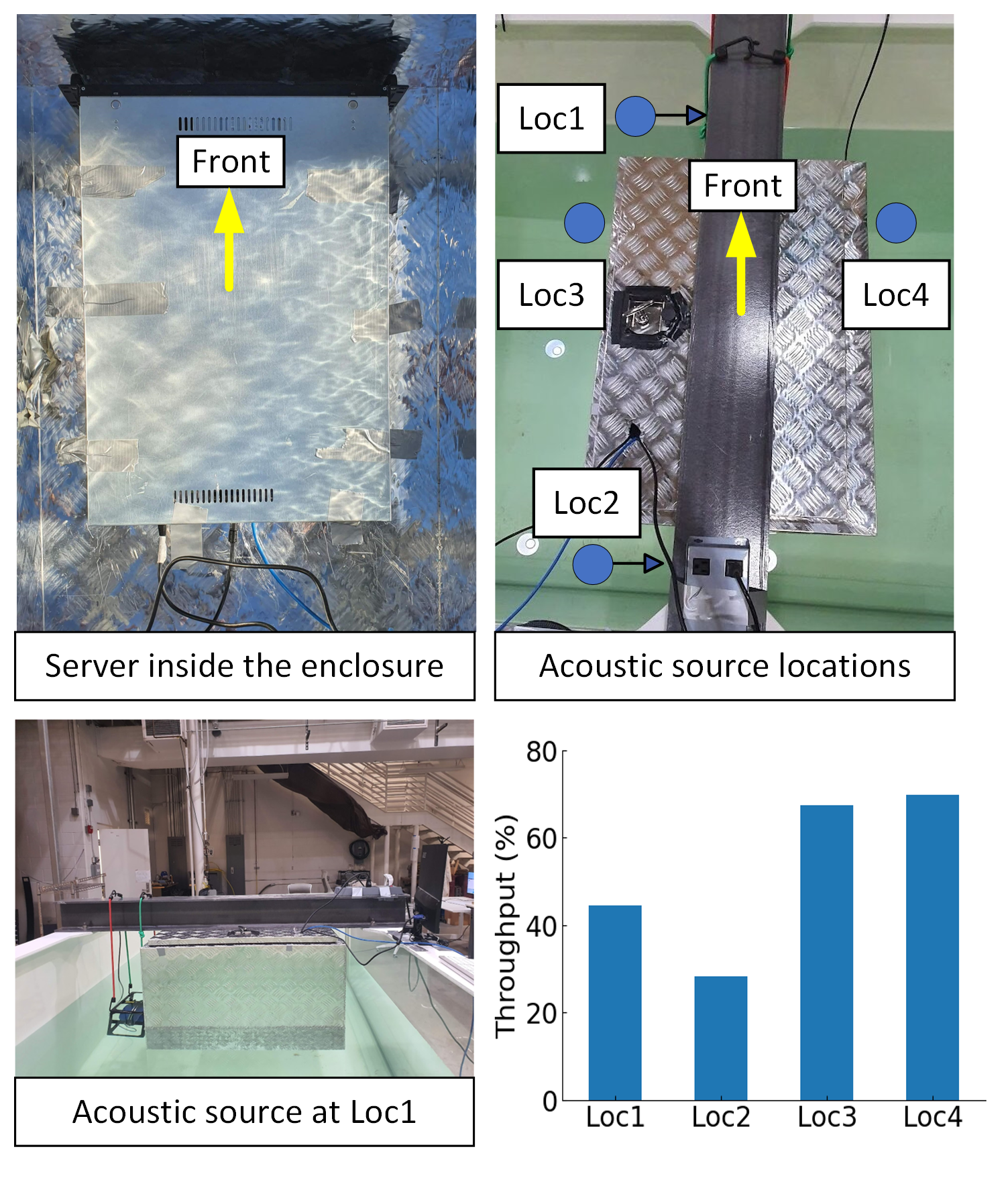}%
\vspace{-2mm}
\caption{Four sound injection points relative to the victim enclosure are shown on top. The bottom left image shows the sound source set up at Location 1. The bar chart shows the normalized RAID 5 throughput during the injection performed at the four injection points.}
\label{fig:positions}
\end{figure}

As described in ~\autoref{sec:theoretical}, sound propagates through mechanical vibration in the rack server and reaches the storage devices. To understand whether this allows the attacker to successfully degrade RAID 5 performance from different injection positions, we measure the throughput of the RAID 5 configuration when running the FIO sequential write benchmark with the sound source placed in different locations around the enclosure containing the target server (see~\autoref{fig:positions}). For each position (described as Locations 1--4 in~\autoref{fig:positions}), we use a fixed volume ($\sim$30 dB $\Delta$SPL) and frequency ($\sim$5.1 kHz, as identified in \autoref{sec:capability:frequency}). As vibrations propagate in the entire structure for their nature, we see that the attacker can cause measurable throughput loss at multiple sound source injection points (including the back of the rack server, far away from the storage device locations in the front). 
While the drop is more severe in some locations than others, the attacker can compensate for this by raising the volume of the injection (see~\autoref{sec:volumeChar}). This reveals how the attacker is not limited to one injection location to pursue the attack.

For the rest of the experiments in this work, we consistently inject sound at Location 1, the front of the rack server, to evaluate the consequences of acoustic injection in a suboptimal location to simulate a realistic scenario.

\subsection{Speaker Orientation}
\begin{figure}[t]
\centering
\includegraphics[width=0.65\columnwidth]{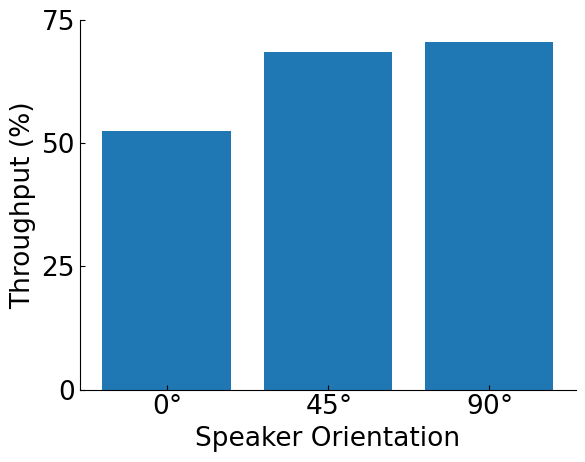}%
\vspace{-1mm}
\caption{FIO measured throughput with the speaker positioned at different angles with respect to the enclosure.}
\label{fig:speakerAngle}
\end{figure}

 To understand if the orientation of the attacker's speaker affects throughput, we measure the RAID 5 throughput with the speaker turned at different angles with respect to the target enclosure (with $0^{\circ}$ representing the speaker aimed towards and $90^{\circ}$ representing the speaker oriented parallel to the target enclosure). In our testbed, we place the speaker 30 cm away from the enclosure to allow for rotation and play the 5.1 kHz tone at 40 dB $\Delta SPL$. We find that the attack is approximately $32 \%$ less effective at $45^{\circ}$ and $34 \%$ less effective at the $90^{\circ}$ angle than in the direct attack case (See Figure~\ref{fig:speakerAngle}). This occurs because the SPL is not uniform at different angles due to the directivity of the speaker, but the lower volume at sub-optimal angles can still cause throughput degradation.

\section{Impact on Critical Operations}
\label{sec:evaluation}
In light of the findings of our characterization analysis, we evaluate the manipulation capabilities of our acoustic injection on popular data center management software and distributed systems.
For this analysis, we use the same setup described in Section~\ref{sec:capability}. In addition, for experiments requiring a second server, we use a PowerEdge R610 rack server~\cite{dellPowerEdge} placed on a table far from the sound source and submerged enclosure. This "on-land" server acts as an unaffected resource for evaluation of data center management software behavior which uses multiple servers such as distributed databases (see~\autoref{sec:cockroachdb}), distributed filesystems (see ~\autoref{sec:HDFS}), and resource allocation managers (see~\autoref{sec:loadManipulation}).

\begin{figure}[t]
\includegraphics[width=\columnwidth]{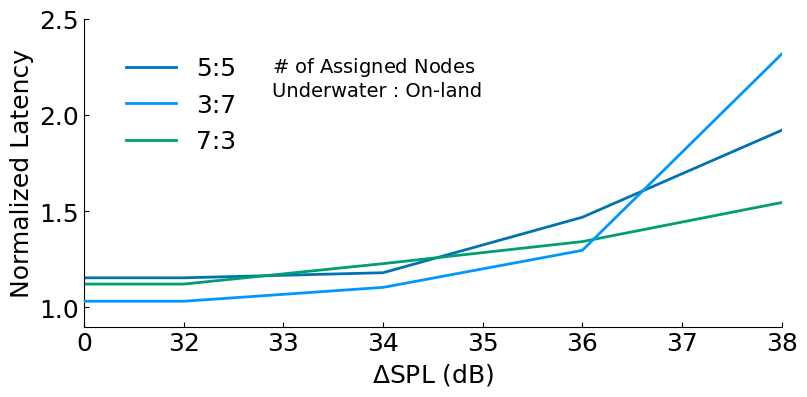}
\caption{Normalized latency for three node assignment configurations when running TPC-C benchmark~\cite{TPCC} on CockroachDB at increasing sound pressure. As the amplitude of the sound injection increases, the latency increases even when the majority of the nodes are assigned to the on-land server.}
\label{fig:cockroachdb}
\end{figure}

\subsection{Latency Control on Distributed Databases}
\label{sec:cockroachdb}
Distributed databases are adopted as a widespread solution to address the need for scalability and high availability, such as in streaming services, and also provide fault tolerance in data centers~\cite{TIDB,VGS2017}.
Compromising distributed databases can lead to service outages~\cite{NETFLIX},  and affect the storage of replicas which in turn can severely degrade the fault tolerance capability of the infrastructure. 

For our analysis, we chose CockroachDB~\cite{TSM2020} as our target to demonstrate the efficacy of our underwater acoustic injection to manipulate a real-world distributed database reliability in terms of latency control.
CockroachDB is a popular, commercially available, and scalable geo-distributed database for high-performance and data processing which has been adopted by many companies, such as Netflix~\cite{NETFLIX} and SpaceX~\cite{SPACEX}.

\vspace{1mm}
\noindent
\textbf{Experimental Setup.} 
We deploy CockroachDB on our aforementioned testbed, which includes two servers, one in the underwater enclosure with RAID 5 configuration while the other on land, outside the influence of underwater acoustic injection. We then consider three different configurations of 10 nodes. In the first configuration, 5 nodes are assigned to the underwater server and the on-land server, respectively. In the second configuration, only 3 nodes are assigned to the underwater server, while in the third configuration, 7 nodes are assigned to the underwater server.

\vspace{1mm}
\noindent
\textbf{Evaluation Metrics.} 
We evaluate the latency variation of CockroachDB when running the TPC-C benchmark~\cite{TPCC}, a transaction processing benchmark provided by CockroachDB as an official performance tester. We adopt the normalized latency as the metric to evaluate the amount of overhead produced by the attack at increasing volumes, which is the measured latency divided by the baseline latency without acoustic injection.
We then inject the sound at the fixed frequency of 5.1 kHz for the benchmark duration, which persists for ten minutes.

\vspace{1mm}
\noindent
\textbf{Results and Observations.}
\autoref{fig:cockroachdb} shows the normalized access latency at increasing injection volumes.
The results show that the performance degradation of CockroachDB~\cite{TSM2020} increases regardless of the number of data nodes. 
Even if fewer nodes are allocated to the underwater server, acoustic injection can almost linearly decrease the overall performance of the CockroachDB cluster with an average latency increase of 43.7\%.
In addition, the underwater attack achieves the highest latency increase at three assigned node settings with 38 dB $\Delta$SPL by 92.7\% on average. 
Once the volume is above 38 dB $\Delta$SPL, the underwater nodes enter an out-of-service state and the operations to the CockroachDB cluster cannot be resolved. This causes abnormal termination of the TPC-C benchmark and removal of the underwater node.

\begin{figure}[t]
\includegraphics[width=0.9\columnwidth]{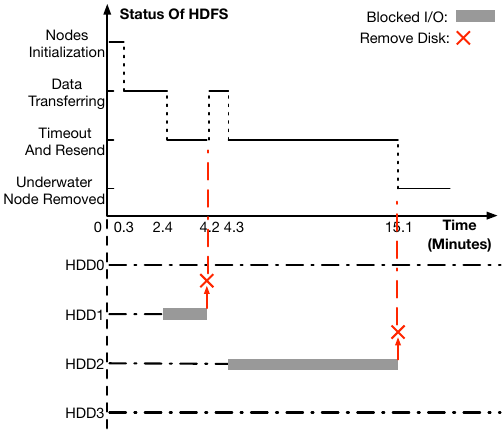}
\caption{The workflow of HDFS with the corresponding liveness of the four HDDs when running the DFSIO benchmark during an acoustic injection at 36 dB $\Delta$SPL.}
\label{fig:hdfs}
\end{figure}

\subsection{Induced Automatic Node Removal in Distributed Filesystems}
\label{sec:HDFS}

Unlike distributed databases which store relational and structured data, distributed filesystems are widely used~\cite{ZCW2021,HADOOP,GGL2003} in data centers for storing unstructured data. 
Latency manipulation in distributed filesystems can provoke severe imbalance of the I/O loads in data nodes and decrease data store reliability because fewer nodes remain available for replica storage.
Thus, we further evaluate our attack capability to manipulate distributed filesystems. Specifically, we evaluate
HDFS~\cite{HDFS}, a popular distributed filesystem which serves as the data store backend of Hadoop~\cite{HADOOP} for high-throughput distributed computing in data centers. 

\vspace{1mm}
\noindent
\textbf{Experimental Setup.}
In this analysis, we adopt the same setup settings used in~\autoref{sec:cockroachdb}, where one HDFS data node is allocated to each server. We then apply our underwater acoustic injection attack at 5.1 kHz frequency at increasing volumes. 

\vspace{1mm}
\noindent
\textbf{Evaluation Metrics.} 
We leverage the Hadoop DFSIO benchmark~\cite{DFSIO} to monitor file accesses to HDFS~\cite{HDFS}. 
We analyze the HDFS logs to extract any changes in workflow and status when accessing the 32 files of 100MB each in the benchmark.
We also monitor the liveness of the storage devices in the RAID 5 during the injection.

\vspace{1mm}
\noindent
\textbf{Results and Observations.}
At volumes below 152 dB SPL, the acoustic injection does not cause any change in status. At 152 dB SPL ($\sim$38 dB $\Delta$SPL), \autoref{fig:hdfs} shows the HDFS workflow and node status while depicting the liveness of each HDD (numbered from 1 to 4) in the RAID 5 configuration. As shown in the graph, the acoustic attack causes abnormal HDD activity which disables the HDFS service after 4.3 minutes.
Specifically, after 2.4 minutes of benchmark execution, the underwater node cannot serve incoming I/O requests due to an HDD that became unresponsive because of the sound injection. 
As a consequence of this effect, the entire RAID 5 became unavailable for data storage.
However, the HDFS stays online for another 1.8 minutes (from 2.4 to 4.2 minutes) as the RAID 5 drops the unresponsive HDD1 due to its unavailability.
When a second HDD became unresponsive (HDD2) at 4.3 minutes for the prolonged attack, the HDFS remains blocked without serving any other file access operation. 
Finally, as RAID 5 requires at least three disks to maintain its correct functioning, and the unresponsive HDDs fall below this threshold, the second unresponsive disk HDD2 is dropped at 15.1 minutes causing the RAID failure. Consequently, the HDFS removes the underwater data node. 

Such automatic removal of the underwater data nodes from the filesystem due to the attack increases the I/O burden on other data nodes and leaves fewer nodes available for storing data replicas dedicated to fault tolerance. 
This shows how an attacker can automatically induce the removal of selective nodes, maliciously redirecting workload over other data nodes causing potential overloading and compromising fault tolerance.

\subsection{Load Manipulation}
\label{sec:loadManipulation}
In this evaluation, we quantify an attacker's ability to manipulate resource monitoring and allocation applications commonly used in data centers. Various organizations, such as Akamai and Cisco~\cite{OpenNebulaCompanies}, use OpenNebula~\cite{milojivcic2011opennebula} to monitor and allocate resources based on server resource availability, and to ensure load balance between nodes~\cite{loadBalancing}. By manipulating resource allocation, an attacker can force tenants to be assigned to slower servers or an already compromised server to enable the other attacks that require tenant colocation~\cite{zhang2012cross, inciseriously, ristenpart2009hey}. Although defenses against colocation attacks have been proposed~\cite{long2020group, agarwal2018co}, they focus on modifying placement policies and do not consider an attacker that can effectively remove a server from the pool of available hardware resources by reducing storage system functionality. 

\vspace{1mm}
\noindent
\textbf{Experimental Setup.}
To evaluate the effect of our attack on resource migration and colocation, we use OpenNebula to balance VM assignments between two servers configured as described in Section~\ref{sec:capability}.
Both servers are connected to a LAN using a switch to a laptop running OpenNebula as an administrator. The administrator laptop monitors the server states and instantiates VMs to be automatically assigned to each server based on resource availability.
Our evaluation is separated into two parts. First, we evaluate how acoustic injection increases latency in individual VMs to determine the affected states. Second, we verify whether an attacker can manipulate the resource manager to force assignment to a particular server (in this case, the attacker forces assignment to the on-land server by blocking the use of the underwater server). This evaluation consists of instantiating 50 VMs and tracking the resource assignments during the acoustic injection at increasing injection volumes.

\subsubsection{Effect on VM Status} In this evaluation, we manually assign and instantiate 3 VMs running an Ubuntu OS and writing 1 GB of data to a file using \textit{dd} on the target server during acoustic injection at increasing volumes. For this experiment, the VMs are assigned to one HDD in the underwater server to observe how HDD vulnerability impacts the VM status. We then constantly increase the volume until the VM experiences disk failure.

\vspace{1mm}
\noindent
\textbf{Evaluation Metrics.} For this experiment, we evaluate the average time taken for each individual VM to complete each state during the acoustic injection. Whenever OpenNebula instantiates a VM, the VM passes through various states (INIT, PROLOG, BOOT, and RUNNING~\cite{OpenNebulaCompanies}). Note that only two of these states (PROLOG and RUNNING) require access to the storage system because the initialization and VM booting run in the application SSD with the server's operating system. The PROLOG state transfers VM files to the host, and, in the RUNNING state, the application runs, meaning that the host server's storage systems will respond to the VM application's I/O requests.

\begin{figure}[t]
\includegraphics[width=\columnwidth]{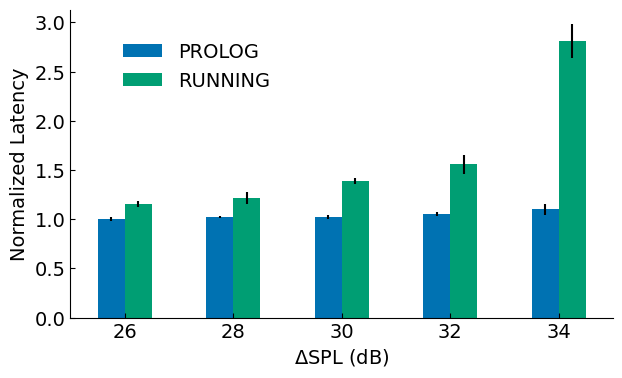}%
\vspace{-1mm}
\caption{Normalized latency increase during the PROLOG and RUNNING states for individual VMs at increasing injection volumes.}
\label{fig:individualVM}
\end{figure}

\vspace{1mm}
\noindent
\textbf{Results and Observations.}
The evaluation results in Figure~\ref{fig:individualVM} shows that
PROLOG and RUNNING VM states have increasing latency with increasing injection volume. For the PROLOG state, the average latency increases up to 10\%, while the average latency for the RUNNING state increases by a maximum of 280\%. The VM fails at 36 dB $\Delta$SPL because the disk becomes unresponsive.

\begin{figure}[t]
\includegraphics[width=\columnwidth]{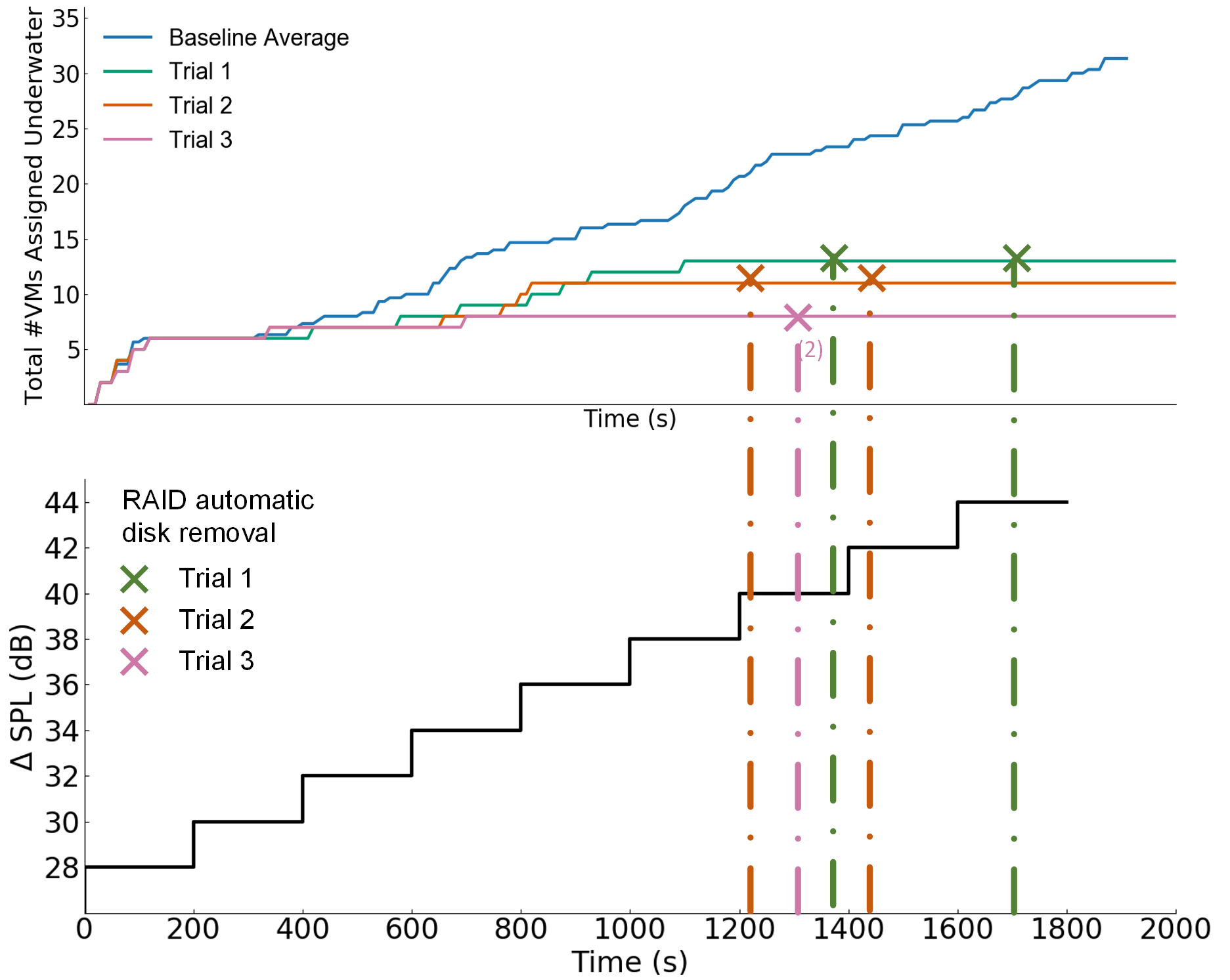}
\caption{(Top) Average number of VMs assigned to the underwater server over time in the no-attack case (blue line) and with a 5.1 kHz acoustic injection. (Bottom) Corresponding volume increment over time. The graphs show the time when disks are automatically dropped from RAID 5. Note that after the second disk drop, the RAID 5 fails.}
\label{fig:VMMigration}
\end{figure}

\subsubsection{Effect on VM Distributions}
For this evaluation, we instantiate 50 VMs
and observe which server OpenNebula automatically assigns each VM to during acoustic injections at increasing injected volumes. In this case, the VMs run on the underwater RAID 5 when assigned to the target server to determine whether the fault tolerance system prevents manipulation of load balancing.

\vspace{1mm}
\noindent
\textbf{Evaluation Metrics.}
We evaluate the number of instantiated VMs assigned to each of the two servers. The attack is considered successful by observing
a 10\% shift in VM assignment from the submerged target to the on-land server. We perform our attack at volumes increasing by 2 dB every 210 seconds, which is 10\% of the average time taken for 50 VMs to finish running with no acoustic injection.

\vspace{1mm}
\noindent
\textbf{Results and Observations.}
Figure~\ref{fig:VMMigration} shows the total VM assignment to the underwater server at increasing volumes. From the results, we can observe a maximum of 74\% and a minimum of 58\% drop in the number of VMs assigned to the underwater server by OpenNebula when reaching up to 44 dB $\Delta$SPL. As in the previous experiments, at high sound levels, the RAID 5 detects the first failure and automatically drops the corresponding disk at approximately 38 dB $\Delta$SPL.
However, RAID 5 continues its operation because it still has 3 of the 4 disks. At 44 dB $\Delta$SPL, RAID 5 drops the second disk causing the RAID 5 to fail, and the VMs become permanently blocked in RUNNING state since they cannot interact with their storage disks. Once the RAID fails, the VMs cannot recover even after the injection stops. 
From these results, we see that the attacker can redirect the VM assignment to the on-land server by decreasing the performance of RAID 5 in the underwater server. We also observe that the overall server performance decreases after each experiment trial, as shown in Figure~\ref{fig:VMMigration}-(top). 
Both disks are dropped sooner from RAID, and the total number of assigned VMs decreases from 13 to 8. This shows how not only the acoustic attack can manipulate load distribution but also induce a permanent degradation of the storage systems even without inducing a complete denial of service.

\subsection{Latency Control on Real-World Workloads}
To understand how acoustic injection can be used to control the latency of real-world workloads, we run the first 50k requests of three SNIA traces~\cite{SNIA} on our submerged server using a RAID 5 device with 4 partitions of 60 GB size each. Among the SNIA traces taken from the operation of real data center workloads for various applications,
we select typical data center workloads of a web server (abbreviated as 'web'), a proxy server (abbreviated as 'prxy'), and a media content server (abbreviated as 'mds'). We run the traces while performing acoustic injection at increasing volumes as in the previous evaluation.

\vspace{1mm}
\noindent
\textbf{Evaluation Metrics.}
We consider the average number of fulfilled I/O requests at increasing volumes for each benchmark. The benchmark is executed using RAIDmeter~\cite{tian2007pro}, a block-level trace replay tool, using a finite, constant timespan for each benchmark based on the time taken to finish sending IO requests ($\sim$38 minutes for mds, $\sim$3 minutes for prxy, and $\sim$22 minutes for web). We characterize attack success as the ability to cause a measurable decrease in fulfilled requests, and we associate the decrease in request fulfillment with the attacker's ability to predictably manipulate application performance.

\begin{figure}[t]
\includegraphics[width=\columnwidth]{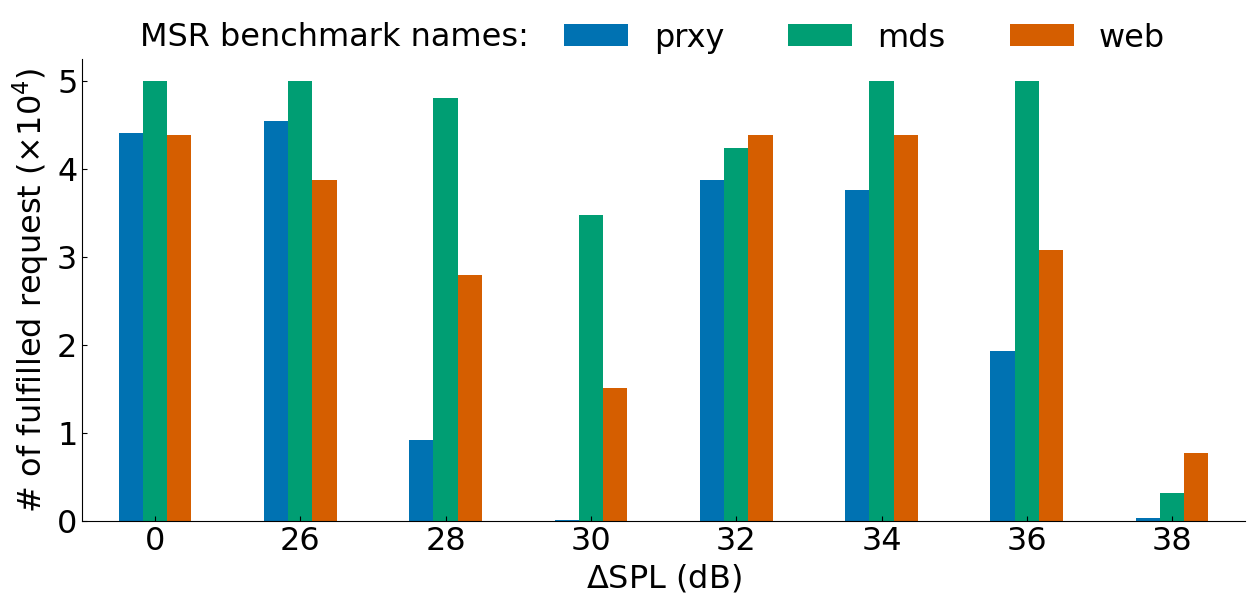}
\caption{Number of fulfilled requests are shown for three MSR benchmarks at increasing injection volumes.}
\label{fig:SNIA}
\end{figure}

\vspace{1mm}
\noindent
\textbf{Results and Observations.}
~\autoref{fig:SNIA} shows each benchmark's normalized request fulfillment results at increasing volumes ranging from 26 to 38 dB $\Delta$SPL. At 40 dB $\Delta$SPL, RAID 5 fails for all trials of all benchmarks. From these results, we see an approximately linear trend with the number of fulfilled requests decreasing in the range from 26 to 30 dB $\Delta$SPL. We note a spike in request fulfillment at 32 dB $\Delta$SPL, which occurs when the slowest disk, which bottlenecks the RAID configuration, is dropped from the array. These results indicate that an acoustic injection can measurably alter the performance of real data center workloads by changing the injection volume.

\subsection{Evaluation on Hybrid Storage Architectures}
\label{sec:hybrid}
Unlike mechanical HDDs, SSDs store data in flash memory on a silicon die and are less likely to be affected by our underwater acoustic injection attack.
As discussed in~\autoref{sec:background:datacenter}, hybrid storage architecture of modern data centers typically deploys SSDs as cache of HDDs~\cite{WLC2020,MJF2010,MLR2014,CWO2011}.
Therefore, we evaluate how our underwater attack affects such a storage architecture.

\vspace{1mm}
\noindent
\textbf{Experimental Setup and Metrics.}
Intel's OpenCAS~\cite{OPENCAS} is a software-level caching tool that allows accelerated access to slow storage devices (e.g., HDDs) by adding a faster device (e.g., SSDs) as a cache.
It is a kernel module that allows the creation of a block device to represent the cached HDDs.
Thus, we leverage OpenCAS to integrate a SSD cache with a write-back policy for HDDs in a RAID 5 configuration using {\em mdadm}.
As in the previous evaluation, we perform the acoustic injection at 5.1 kHz with a fixed volume of 30 dB $\Delta$SPL based on the previous observations.
Then, we use FIO to evaluate the performance of the cached HDDs to evaluate the attacker's capabilities through monitoring the latency and bandwidth of four selected workloads: sequential write (SW), sequential read (SR), random write (RW), and random read (RR).
Moreover, we vary the allocated SSD size of the cache to demonstrate how the cache size impacts the efficiency of our attack.

\begin{figure}[t]
	\centering
	\subfloat[Cache size $0.5$ GB. \label{rwcdf0.5}]{%
		
			\includegraphics[width=0.47\columnwidth]{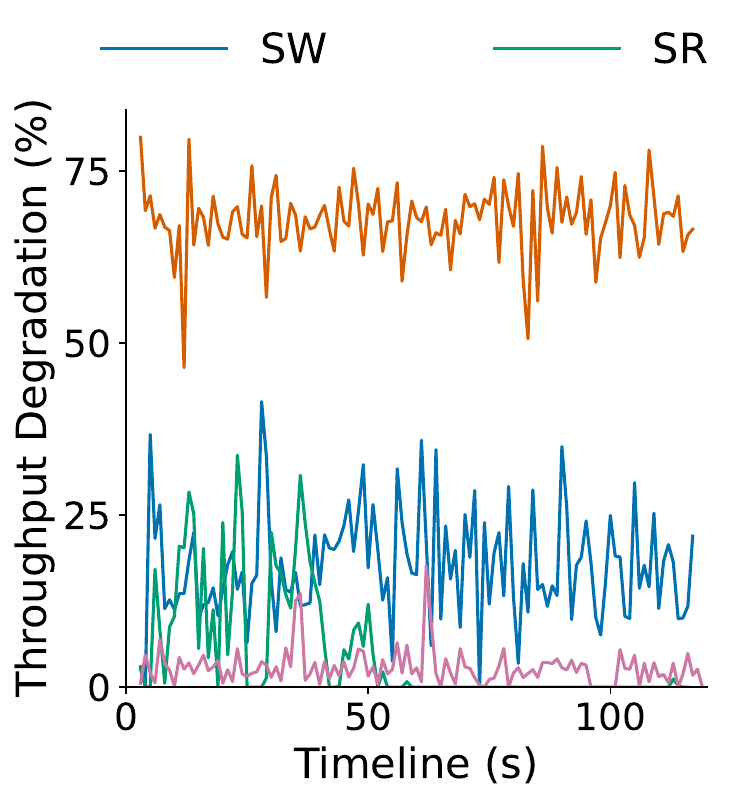}
		}
	\hspace{\fill}
	\subfloat[Cache size $1$ GB. \label{rwcdf1}]{%
		\includegraphics[width=0.47\columnwidth]{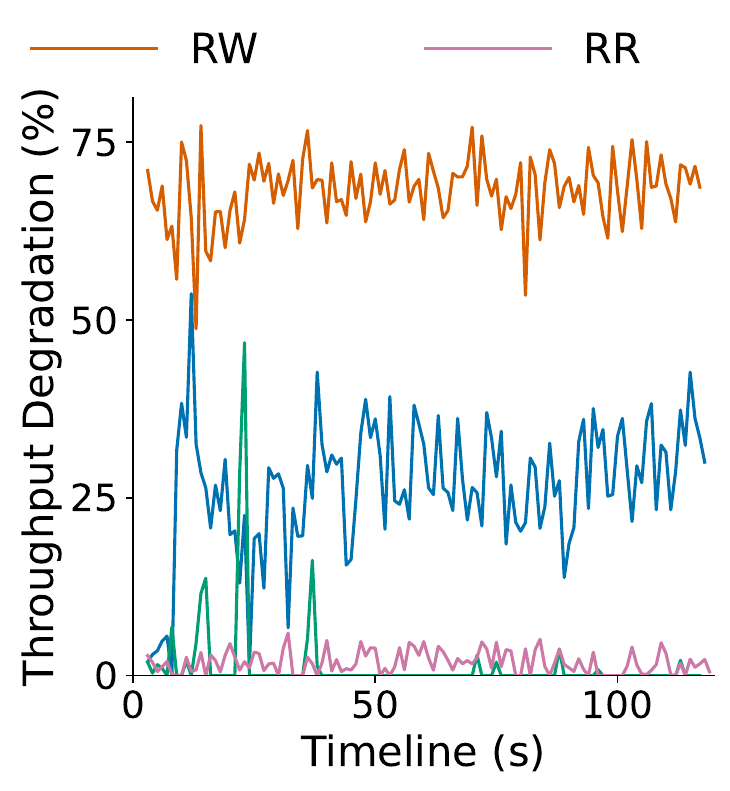}
	}

	\subfloat[Cache size $1.5$ GB. \label{rwcdf1.5}]{%
		\includegraphics[width=0.47\columnwidth]{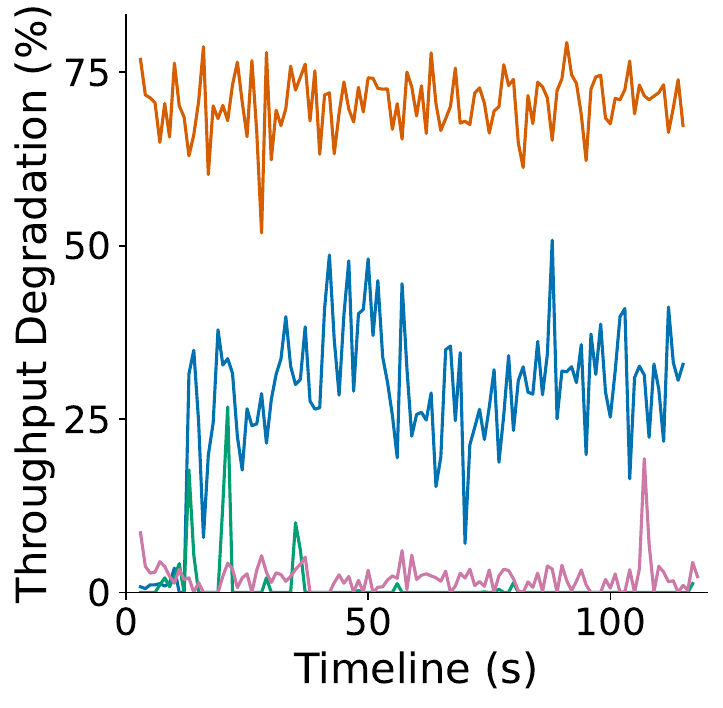}
	}
	\hspace{\fill}
	\subfloat[Cache size $2$ GB. \label{rwcdf2}]{%
		\includegraphics[width=0.47\columnwidth]{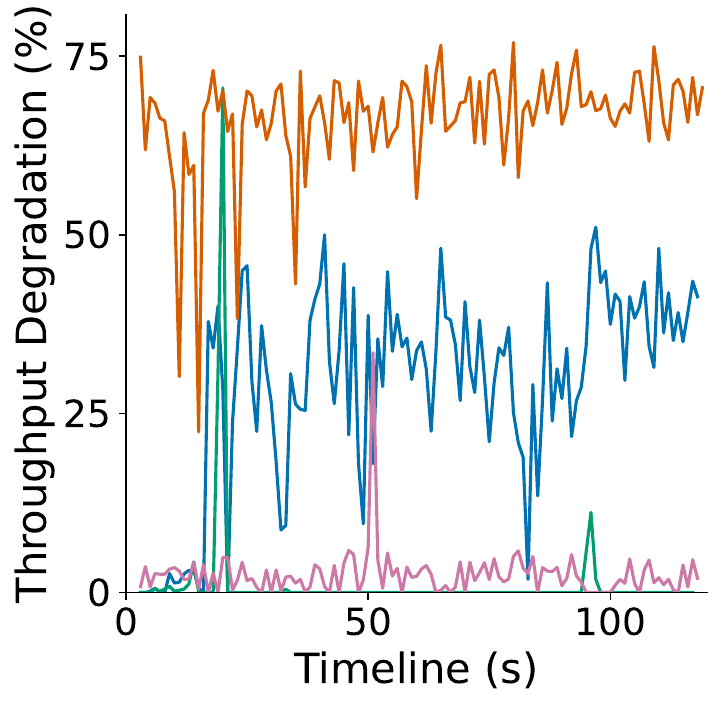}
	}\\
	\caption{Bandwidth degradation percentage incurred by our underwater attack during the running of sequential write (SW), sequential read (SR), random write (RW), and random read (RR) workloads of FIO.}	\label{fig:bw_change}
\end{figure}

\begin{figure}[t]
	\centering
	\subfloat[Cache size $0.5$ GB. \label{2rwcdf0.5}]{%
		
			\includegraphics[width=0.47\columnwidth]{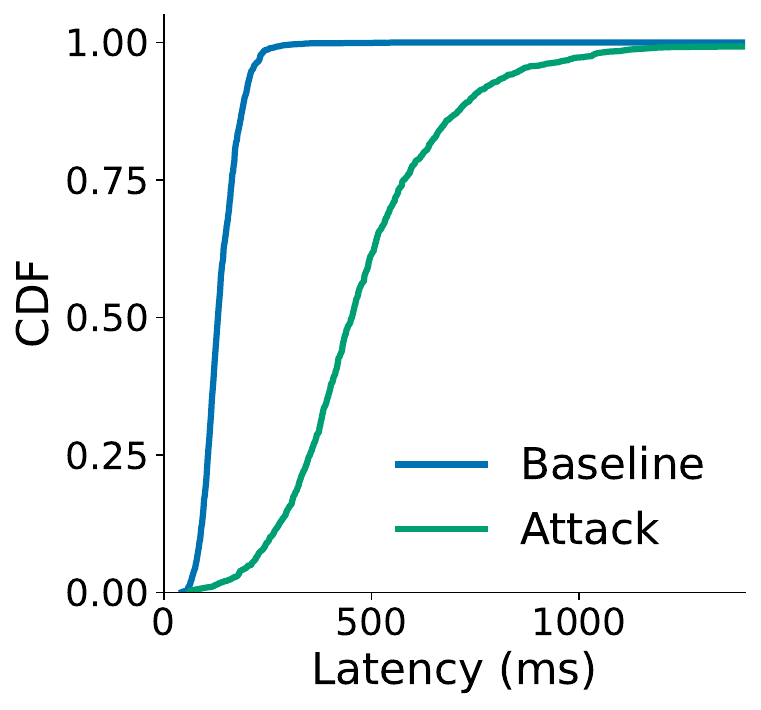}
		}
	\hspace{\fill}
	\subfloat[Cache size $1$ GB. \label{2rwcdf1}]{%
		\includegraphics[width=0.47\columnwidth]{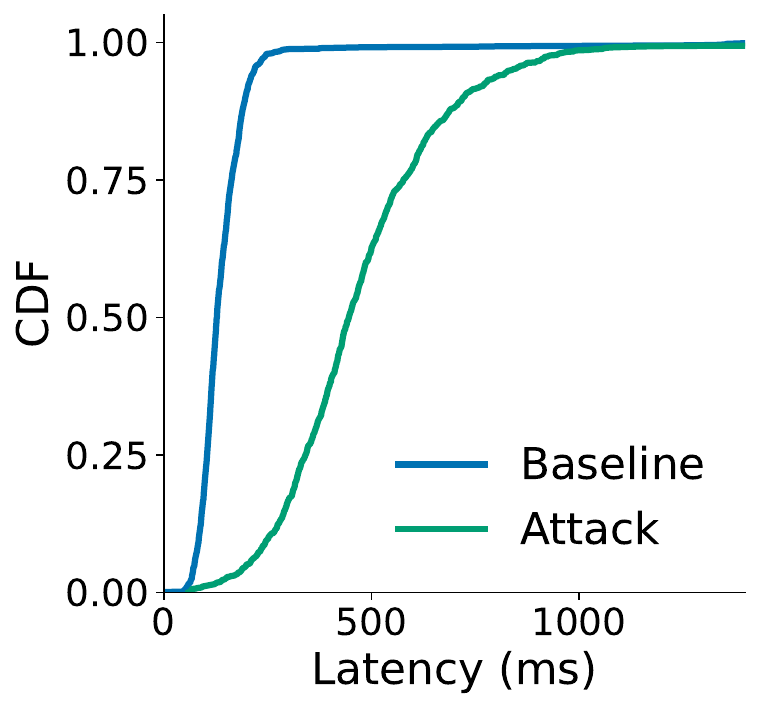}
	}

	\subfloat[Cache size $1.5$ GB. \label{2rwcdf1.5}]{%
		\includegraphics[width=0.47\columnwidth]{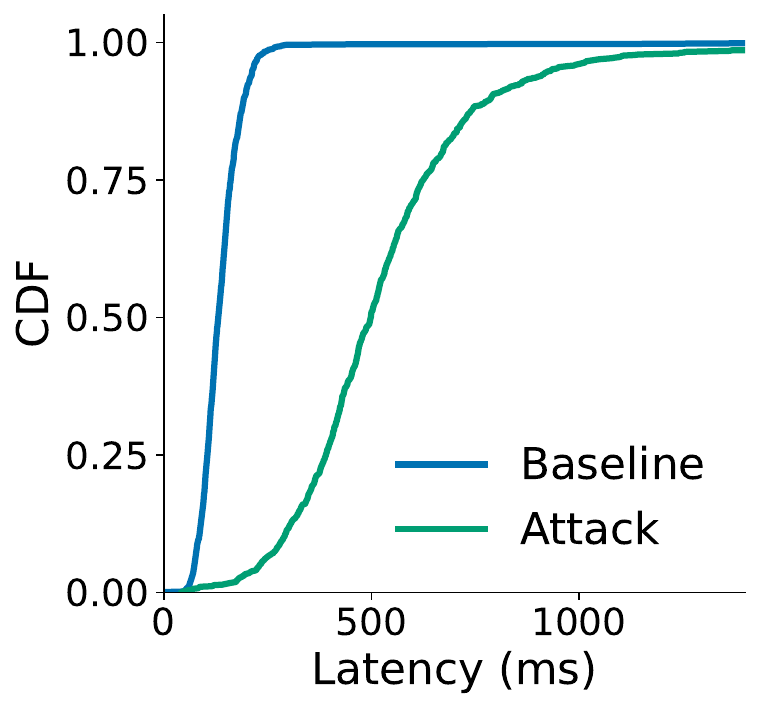}
	}
	\hspace{\fill}
	\subfloat[Cache size $2$ GB. \label{2rwcdf2}]{%
		\includegraphics[width=0.47\columnwidth]{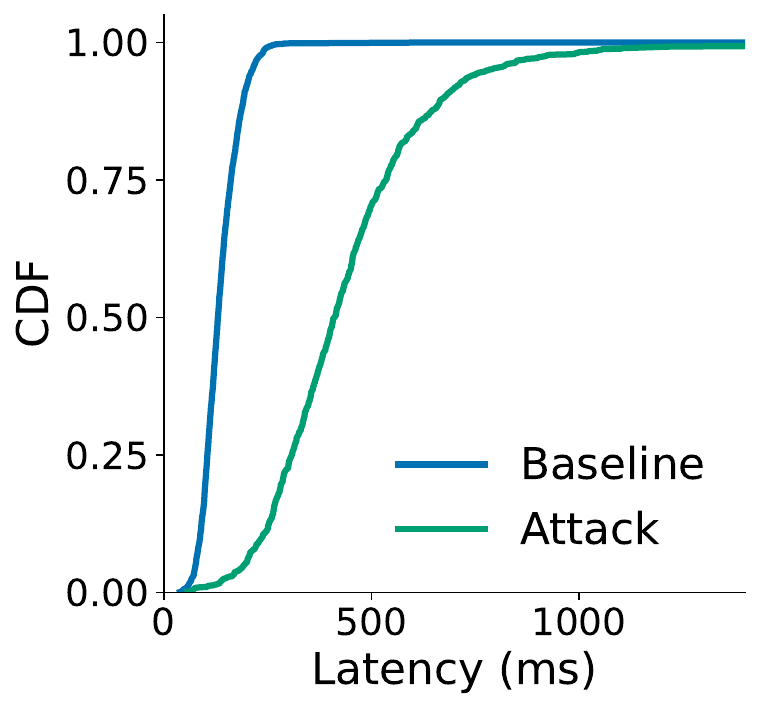}
	}\\
	\caption{CDF of latency when running the {random write} (RW) workload of FIO with or without underwater injection.}
	\label{fig:cdf_rw}
\end{figure}

\begin{figure}[t]
	\centering
	\subfloat[Cache size $0.5$ GB. \label{swcdf0.5}]{%
		
			\includegraphics[width=0.47\columnwidth]{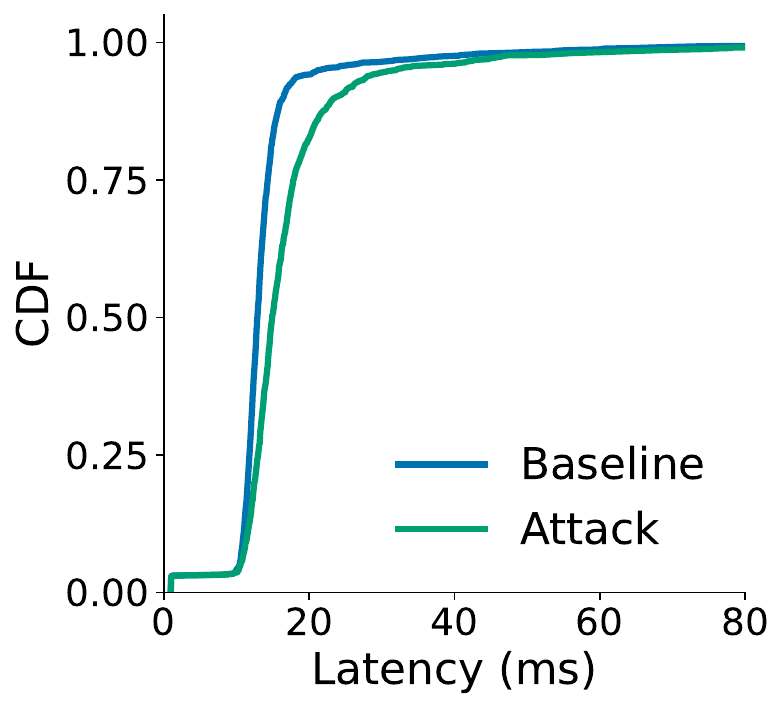}
		}
	\hspace{\fill}
	\subfloat[Cache size 1GB. \label{swcdf1}]{%
		\includegraphics[width=0.47\columnwidth]{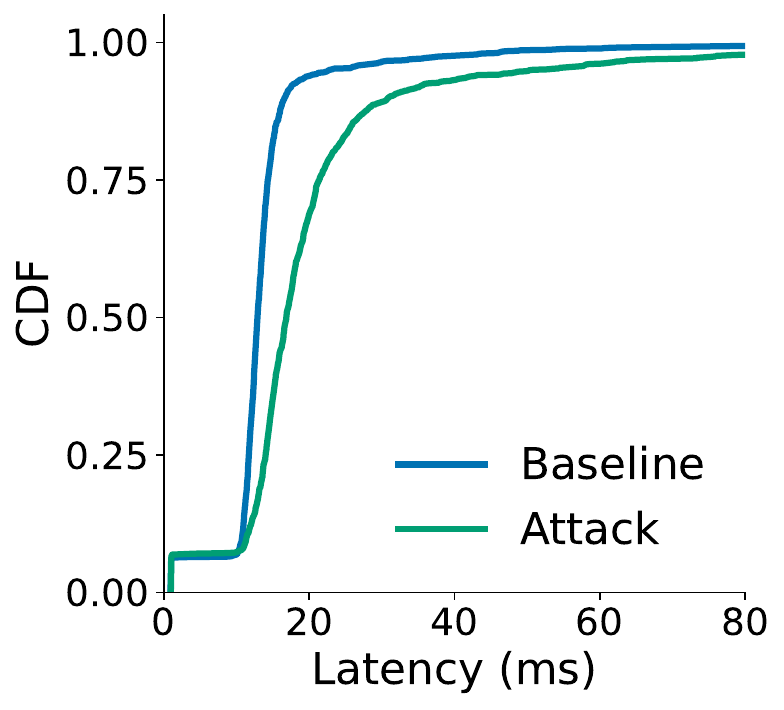}
	}

	\subfloat[Cache size $1.5$ GB. \label{swcdf1.5}]{%
		\includegraphics[width=0.47\columnwidth]{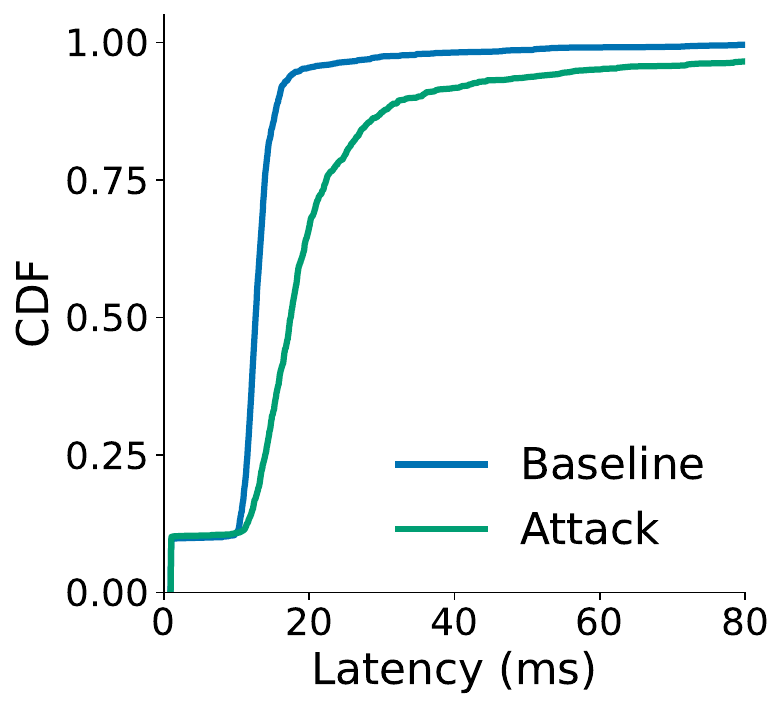}
	}
	\hspace{\fill}
	\subfloat[Cache size $2$ GB. \label{swcdf2}]{%
		\includegraphics[width=0.47\columnwidth]{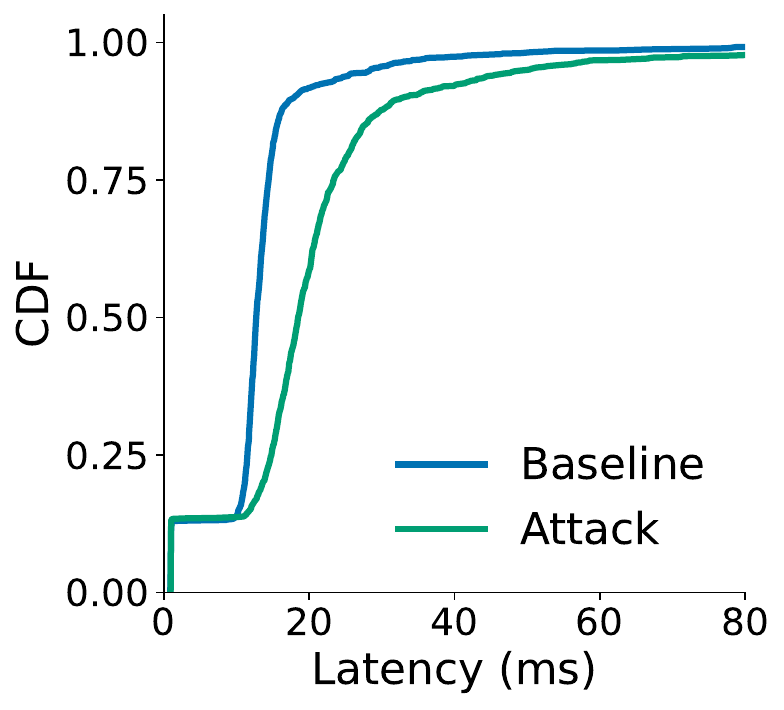}
	}\\
	\caption{CDF of latency when running the {sequential write} (SW) workload of FIO with or without underwater injection.}
	\label{fig:cdf_sw}
\end{figure}

\vspace{1mm}
\noindent
\textbf{Results and Observations.} 
\autoref{fig:bw_change} shows the bandwidth degradation when running the four FIO workloads with different cache sizes.
As observed in the previous experiments in~\autoref{sec:capability:frequency} write operations are more affected by the acoustic injection leading to a more noticeable performance degradation in write-intensive workloads than in read-intensive workloads.

For workloads with more random access behaviors (i.e., RW and RR), the performance degradation is more significant because random data access to HDDs will incur frequent HDD actuator arm movement, making it easier to be affected by the sound-induced vibrations.
Moreover, the hit ratio of RW workload is under 1\%, whereas sequential write achieves a higher hit ratio, where its hit ratios are 33.3\%, 56.9\%, 68.6\%, and 76.1\% when allocating 0.5 GB, 1 GB, 1.5 GB, and 2 GB cache size respectively.
Therefore, the bandwidth degradation provoked by the attack is alleviated in workloads with a high hit ratio.
Figures \ref{fig:cdf_rw} and\ref{fig:cdf_sw} show the cumulative distribution function (CDF) of access latency at different cache sizes when running RW and SW workloads in our attack.
In the RW workload, the access latency under the attack always distributes between 200--800 ms, while the access latency fits within 1--200 ms in the benign case. 
Even if the cache size increases, the latency increase incurred by our attack is still significant.
Since the RW workload presents a low hit ratio -- less than 1\% -- to the cache, our attack can significantly degrade the performance of the cached HDDs because most of the I/O requests are served by the HDDs.
In contrast, a large amount of I/O requests are served by the cache in the SW workload, thus our attack has less impact on sequential writes. 
However, the performance degradation is still nontrivial, as shown in~\autoref{fig:cdf_rw} and~\autoref{fig:cdf_sw}.
Therefore, our underwater acoustic injection attack can make the storage system unpredictable, which is critical to provide deterministic latencies~\cite{HTL2020,LZL2019} desired by data center providers, with overall performance degradation.
\subsection{Evaluation for Open-Water Deployments}
\label{sec:evaluation:openwater}
To evaluate whether such acoustic attacks can be performed in open water and to understand the distance limit for the attacker, we deploy our testbed setup to a lake (see Figure~\ref{fig:overview}c). For this scenario, we weighed the metal enclosure with bags of sand to reach the required water level. Then, we measure the RAID 5 write throughput at increasing volumes and distances from the sound source.

\vspace{1mm}
\noindent
\textbf{Evaluation Metrics.}
We use FIO~\cite{FIO} to record RAID 5 throughput over 30-second spans for 3 consecutive trials. For our volume variation evaluation, we consider a 30 cm distance from the sound source. For our distance evaluation, we consider the maximum achievable distance where the attack can successfully degrade the RAID 5 performance.

\vspace{1mm}
\noindent
\textbf{Results and Observations.} Figure~\ref{fig:volumes} (b) shows the throughput variation at increasing volumes. We observe a similar degradation as in the laboratory testbed scenario (Figure~\ref{fig:volumes} (a)), but a higher volume is required to reach the same amount of degradation. We suspect that the bags of sand altered the propagation properties of the vibrations. Such results also indicate how our laboratory setup, even if limited, can be used to simulate realistic scenarios. 

For the maximum achievable distance, we induce an average degradation of 61\% at 6.35 meters from the enclosure, which represents the maximum distance available in our lake scenario. This result shows how sophisticated acoustic injection attacks can be performed at far distances with commercially available speakers.

\subsection{Finite Element Simulation}
\label{sec:evaluation:simulation}

At the time of writing, there are no available testing facilities for commercial UDC deployments, thus in our testbed evaluation, we approximate the UDC vessel with an aluminum enclosure.
To provide a more realistic preliminary analysis of the attack, we simulate our acoustic injection using a COMSOL Finite Element Method (FEM)~\cite{multiphisics2014comsol} model. As used in previous work~\cite{bolton2018blue}, such modelling allows combining multiple physics phenomena for simulations of real-world scenarios, such as, the sound propagation between two media (seawater and the vessel mechanical structure).

\begin{figure}[t!]
\includegraphics[width=\columnwidth]{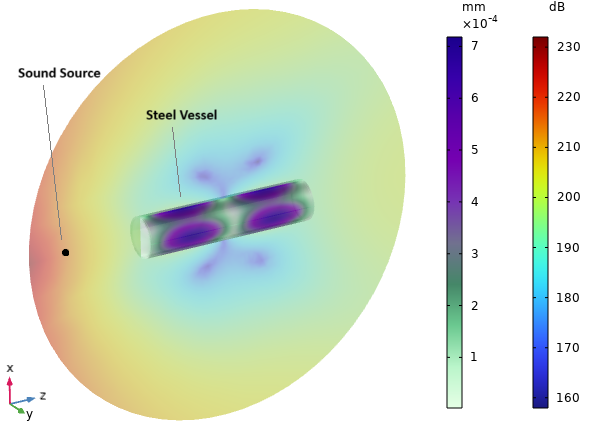}
\caption{COMSOL simulation of the pressure and displacement of the 1/100 scale model of the Project Natick vessel~\cite{dataCenterDimensions} under a 220 dB SPL sound source in seawater.}
\label{fig:comsolPressureWaves}
\end{figure}

We based our analysis on Microsoft’s Project Natick resources~\cite{microsoftunderwaterdatacenterarticle} and publicly available information on subsea vessel prototypes. 
We build a 1/100 scaled steel hollow vessel (12.2 m x 1.4 m radius)~\cite{dataCenterDimensions}) with 11.7 mm steel thickness based on thickness recommendations for underwater pressure vessels~\cite{pressureVesselThickness} (See Figure~\ref{fig:comsolPressureWaves}). We scale the model to allow for a finer-grained mesh for more accurate simulation results.  We account for the 35 m depth below sea level as described for Project Natick, with a salinity level of 35 (reference salinity for seawater~\cite{intergovernmental2010international}). We consider a budgeted attacker with a military-grade speaker which can reach SPL of 220 dB (based on the SPL of sonars~\cite{soundComparison}) simulated in our model as a sound source generating spherical waves facing the flat surface of the vessel as depicted in Figure~\ref{fig:comsolPressureWaves}.  As described in our theoretical analysis in Section~\ref{sec:theoretical}, such vibrations propagate to the internal server racks and storage devices through their contact surfaces. 

After simulating a frequency sweep of eigenfrequencies to find the structure resonance frequency, we set 6.95 KHz as our injection frequency. We then simulate the injection at 6 cm from the structure as per our capability evaluation, achieving an average total displacement of $\sim$360 nm along the three orientation axis with a maximum of $\sim$718 nm. 
We then estimate the maximum capability to induce mechanical vibrations on the full-scale vessel, measured in terms of structure displacement along the three orientation axes, reaching an average total displacement of $\sim$145.5 nm.
As a reference, per our PES analysis in Section~\ref{sec:PES} and the literature~\cite{kwong2019harddriveofhearing, bolton2018blue} typically hard disk read and write from the magnetic platters by the read/write head, which floats about $\sim$5 nm above the disk surface, and, in the case of enterprise-range drives used in data centers, can deviate from the center of the track by no more than $\sim$7 nm to avoid reading and writing errors. 
Such simulation results indicate how acoustic injections can potentially generate vibrations strong enough to propagate inside steel vessel structures. 
Based on Eqs.~\ref{eq:attenuation} and  \ref{eq:solidLiquidBoundary}, we know that the SPL in seawater attenuates exponentially, and the displacement induced in a solid structure is proportional to the applied force given by the intensity of the injected sound. Therefore, we can estimate the maximum distance achievable by our model using a conservative attenuation coefficient $\alpha$ of $10^{-1}$ Nepers/km ~\cite{ainslie2010propagation} (this value is taken at 10 kHz reference frequency while our frequency is lower). We find that such an attacker can theoretically induce an average of 131.2 nm displacement at 1 km from the structure, revealing small decrease in vibration over large distances. 

\section{Defenses}
\subsection{Potential Defenses}
\begin{figure}[t]
\includegraphics[width=\columnwidth]{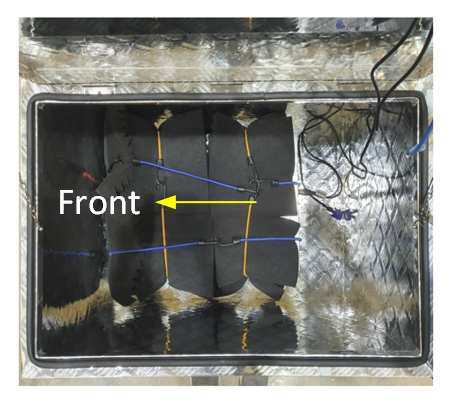}%
\vspace{-3mm}
\caption{Server wrapped in sawtooth-shaped sound-
absorption panels to test the effectiveness of passive attenuation against our acoustic injection attack.}
\label{fig:foamBox}
\end{figure}

\noindent
\textbf{Passive Attenuation Using Absorptive Material.}
\label{sec:foam}
One commonly suggested defense against acoustic injection attacks is the use of sound absorption materials to passively attenuate sound-induced vibrations. However, Blue Note~\cite{bolton2018blue} demonstrates that this measure does not block degradation caused by lower-frequency injections and increases HDD operational temperature in the air, which increases the risk of drive failure. To test the limitations of passive attenuation on our submerged setup, we use sawtooth-shaped sound-absorption panels to wrap our submerged server (see~\autoref{fig:foamBox}). We repeat the experiment in ~\autoref{sec:volumeChar} at the highest volume that our speaker can achieve (180 dB SPL).

\begin{figure}[t]
\includegraphics[width=\columnwidth]{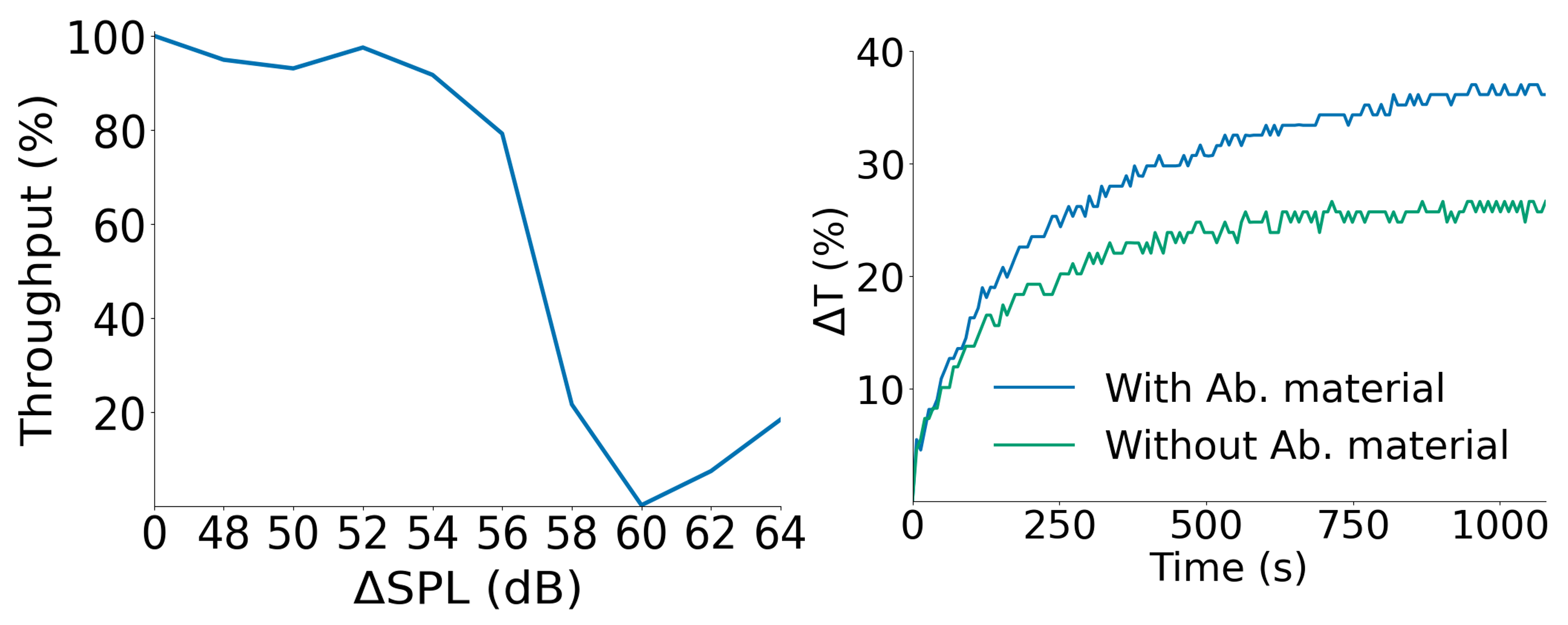}
\caption{(Left) RAID 5 write throughput at increasing injection volumes at 5.1 kHz injection frequency. Note that in 2 of 3 trials, a disk was automatically removed from RAID 5 at 60 dB $\Delta$SPL above background noise. This causes the increasing throughput at 62 and 64 dB $\Delta$SPL. (Right) Temperature increases in the presence of the absorbing material.}
\label{fig:foamSetup}
\end{figure}

Our results in~\autoref{fig:foamSetup} show that we can cause similar changes in throughput, meaning that an attacker could overpower the vibration absorption by increasing the sound volume. We also evaluate the temperature of the server with and without the absorbing material by running a CPU stress test using the \textit{stress} utility~\cite{stress} for 20 minutes and logging the average temperature of the server's CPU cores. From the results, we see about a 10\% difference in the temperature increase with and without the foam for a single server. Microsoft's Project Natick submerged data center contains 864 servers~\cite{microsoftunderwaterdatacenterarticle} that generate heat in an enclosed space; this dense configuration implies a significant increase in heat retention, which will be unsustainable for server health. Our results show that the use of sound absorption materials requires a careful redesign of the internal data center structure by considering the tradeoff between cooling efficiency and acoustic attack protection. Design solutions proposed in research for acoustic attenuation typically focus on attenuating internal fan and disk noise~\cite{killeen2023fan, wasala2022acoustic, joshi2012introduction} rather than mechanical vibrations coming from external sources. Furthermore, they might require the isolation of each individual server rack in materials such as polyurethane acoustic foam (as the one used in our experiments) or custom acoustic metamaterials targeting specific frequency ranges~\cite{ramamoorthy2018towards}. These solutions might be adapted to target the resonance frequency ranges exploited by the attacker, leveraging the data center's internal server configuration.

\vspace{1mm}
\noindent
\textbf{Active Noise Cancellation.}
Another commonly suggested defense against acoustic injection attacks is noise cancellation. Previous works have argued that noise cancellation is an impractical defense because it is difficult to generate a noise signal with the required equal amplitude to the injected signal and which envelops the entire region~\cite{bolton2018blue}. Another consideration in the underwater scenario is noise pollution. As described in~\autoref{sec:background:waterprops}, sound travels faster in water than in air, so emitting high-volume sound waves surrounding the data center could be detrimental to the environment. Marine life, such as whales~\cite{weilgart2007impacts} and fishes~\cite{weilgart2018impact}, is harmed by noise pollution. As such, this defensive measure would affect the overall environmental sustainability of underwater datacenters. In addition, acoustic emission can interfere with sound-based communication systems underwater.

\vspace{1mm}
\noindent
\textbf{Sensor Fusion for Detection.}
Sensor fusion-based detection techniques using hydrophones, accelerometers, and vibration sensors could also be used to detect acoustic injection attacks on data centers. Such defensive measures imply the deployment of additional hardware and detection control software to the data center. However, it is worth noticing that external sensors such as cameras, accelerometers, and microphones are also vulnerable to the effect of acoustic vibrations at the resonance frequency which can cause measurement errors, false triggering, and DoS, as demonstrated in previous works~\cite{trippel2017walnut, ji2021poltergeist, kune2013ghost, sugawara2020light}.

\vspace{1mm}
\noindent
\textbf{Feedback Controller and Firmware Modifications.}
HDDs have feedback controllers that compensate for vibrations in a narrow band of frequencies to prevent disruption~\cite{teoh2008rejecting}. For instance, Bolton~et~al.~\cite{bolton2018blue} implemented in simulation an augmented feedback controller by updating the storage device firmware. This and other hard disk proprietary firmware modifications can be applied to attenuate vibrations up to a certain displacement level on single hard disk drives.

\vspace{1mm}
\noindent
\textbf{Physical Security.}
Our open-water scenario shows how our attack can be deployed more than 6 meters away from the targeted enclosure with the maximum volume achievable by our setup (commercial speaker and amplifier). In UDC deployments, physical surveillance mechanisms such as high-resolution underwater cameras and motion sensors controlled by trained personnel can be placed to prevent attackers from reaching the enclosure's proximity to perform the attack. The detection range depends on their visibility range and resolution which might vary based on the depth and light attenuation~\cite{zhou2023underwater} reaching a maximum of 20-30 meters (approximately 65-100 feet) in clear water~\cite{kulshreshtha2017estimation}. Our simulation based on the Project Natick~\cite{microsoftunderwaterdatacenterarticle} prototype deployment shows significant vibration displacement at 1 km from the structure using a 220 dB SPL sound source similar military-grade sonars used in the real world~\cite{soundComparison}. This preliminary analysis reveals that physical security mechanisms should be designed to take into account the attacker's capabilities, surveillance sensor accuracy, and environmental conditions.

\subsection{Proposed Defense}
\vspace{-2mm}
We propose a proof-of-concept detection mechanism that uses a machine learning model to detect multiple simultaneous, low-volume throughput degradations instead of attenuating single disk vibrations. Our defense relies on analyzing the throughput of disk clusters in close physical proximity to differentiate between normal performance degradation and acoustic injection attacks. This approach is based on the idea that sound-induced vibrations affect multiple disks simultaneously because sound radiates, generating a pattern of throughput changes that can be detected. Such a defense can be deployed at the cloud resource management level without requiring access to proprietary HDD firmware or datacenter physical structure redesign.

\vspace{1mm}
\noindent
\textbf{Evaluation Method and Results.} To identify throughput degradation in multiple disks, we consider the full-HDD architecture and generate a profile of each of the four disks in the RAID 5 configuration. Such profiling can be customized based on the system, and for our proof-of-concept analysis, we use the FIO~\cite{FIO} sequential write workload for 30 seconds. We first collect the storage system throughput for 100 trials on a 100 MB partition for each disk without any acoustic injection.
We then run 100 30-second trials of the same benchmark during acoustic injection at 26 dB $\Delta$SPL, 28 dB $\Delta$SPL, and 30 dB $\Delta$SPL, the lowest volumes which cause the minimal throughput change in our scenarios. To detect the attack for different injection volumes, we use k-means clustering with Partial Curve Mapping (PCM)~\cite{witowski2012parameter} metric. PCM uses arch length and area of the throughput data with respect to time to measure the similarity between disk performances. To quantify our defense's ability to differentiate between attack and no-attack cases, we generate 1,000 combinations of all four disks with random benign and attack throughputs. We repeat this for each volume level. We consider an attack if at least three disks show anomalous throughput behavior. Through this evaluation, we achieve a 0\% False Positive Rate and 98.2\% True Positive Rate. Although we only consider a limited set of hard disk drives and FIO benchmark profiling, this proof-of-concept defense shows how the use of ML techniques can allow the recognition of localized degradation patterns that can reveal the presence of potential acoustic attacks.

\vspace{1mm}
\noindent
\textbf{Post-detection Defense.}
Recent replication~\cite{CEK2015,LSC2021} and erasure coding~\cite{HSX2012,HCY2021,CDL2014} optimization techniques explored in research can provide selective data redundancy for preventing data loss and ensuring high-quality fast data recovery in cloud settings. Replication techniques generate replicas of data and distribute them to multiple storage nodes located at different places, while erasure coding techniques compute multiple parities for stored data and use the erasure coding calculation with stored parities to recover failed storage nodes. Upon detecting an attack, the resource management system can be configured to leverage these advanced techniques to migrate the I/O requests to specific unaffected nodes outside the realm of the sound-affected areas which house the replicas of the affected data. This is possible since, as shown in our empirical evaluation, the storage regions affected by our attack are physically adjacent to each other due to the nature of the vulnerability.
\section{Discussion}
\noindent
\textbf{Long-Term Disk Degradation.}
Through our experiments, we note that hard disks suffer from long-term degradation due to acoustic injection attacks. While disks can be re-added to RAID and undetected disks can sometimes be re-detected by rebooting or by physical reconnection, three HDDs became completely undetectable and permanently damaged during our experiments. In underwater infrastructure deployments where the ability to access enclosed vessels to replace unresponsive or damaged storage devices is limited, even short-lived acoustic injections can potentially cause severe performance loss. 
For instance, the latencies observed before and after the injections for the MSR benchmark of a proxy server ({\em prxy}) showed a $\sim$1,350\% increase in average request latency. Thus, the attack continues to have an effect after the end of the injection.

\vspace{1mm}
\noindent
\textbf{RAID 5 Disk Bottleneck.}
Throughout our evaluations, we found a sudden increase in throughput after a regular decrease in throughput (see~\autoref{fig:foamSetup},~\autoref{fig:volumes}, and~\autoref{fig:SNIA}). This sudden spike coincided with the first RAID 5 automatically removing the slowest disk from the configuration (we used a 4-disk RAID 5 configuration, and RAID 5 requires a minimum of 3 disks). RAID 5 write requests require that parity is written to all disks in the configuration~\cite{RAID5}, so dropping the slowest disk can partially alleviate the bottleneck on parity writing. However, this effect rapidly decreases with increasing injection volumes.

\vspace{1mm}
\noindent
\textbf{Hybrid Storage Architecture.}
Existing data centers employ SSDs as the cache for HDDs because of limitations in the SSD technology as discussed in~\autoref{sec:background:datacenter}.
Since SSDs contain no mechanical component, they are immune to performance degradation caused by acoustic vibrations~\cite{bolton2018blue}, making them a potential solution to alleviate our underwater attacks.
However, the cache cannot guarantee that data will not be evicted from SSDs to HDDs, and SSDs can be overused, which motivates redirect write requests to HDDs~\cite{WLC2020}.
Therefore, acoustic attack remains a security threat to underwater infrastructure, as reliance on HDDs remains widespread.

\vspace{1mm}
\noindent
\textbf{Limitations.}
In this study, we evaluate an attacker's ability to reduce RAID 5 device performance using acoustic injection. Our laboratory testbed is an imperfect approximation of a submerged data center, which would be the real-world target of such attackers.
While we use a FEM model to simulate a prototype deployment under more realistic attack scenarios, the simulation does not fully capture all the factors of a real-world subsea environment.
Furthermore, real-world UDCs might significantly differ in size, type of enclosure, deployment in rough or salty water, and other physical parameters that might considerably impact the results of audio injection.
However, our analysis, even if limited to proof-of-concept scenarios, unveils new sophisticated attack vectors that go beyond the simple Denial-of-Service, influencing the behavior and reliability of traditional fault tolerance and load-balancing storage techniques that cannot withstand acoustic attacks.
We also do not consider different RAID configurations or non-cache-based SSD hybrid architectures. Our analysis is limited to one single server deployment underwater and the application analysis focuses on standard benchmarks of realistic data center workflows (e.g., SNIA traces). Formal analysis has been used in previous work~\cite{BGB2018, liu2022bridging} to validate the design of cloud storage systems. However, none of the existing works has applied formal analysis for a large-scale study of the throughput degradation incurred by acoustic attacks in cloud nodes. We leave this analysis as future work.

\vspace{1mm}
\noindent
\textbf{Acoustic Safety Considerations.}
All the experiments in this work have been conducted in controlled isolated environments with participants wearing the appropriate hearing protections.

\section{Related Work}
\noindent
\textbf{Signal Injection Attacks.}
Signal injection attacks have been performed using a wide range of signal types, including optical~\cite{sugawara2020light, cao2023you}, acoustic~\cite{trippel2017walnut, tu2018injected}, and electromagnetic waves~\cite{tu2019trick, wang2022ghosttouch, kune2013ghost}. By exploiting component sensitivity to these signal types, researchers have performed attacks on various devices, including temperature sensors~\cite{tu2019trick}, hard disk drives~\cite{bolton2018blue, sheldon2023deep, shahrad2018acoustic}, autonomous vehicles~\cite{cao2023you}, underwater acoustic networks~\cite{zuba2011launching, zuba2015vulnerabilities, dong2014study, xiao2015experimental}, and automatic speech recognition systems~\cite{sugawara2020light,zhang2017dolphinattack}. Unlike these previous works, we investigate acoustic injection in fault-tolerant storage configurations and data center management systems, which are not designed to process acoustic signals.

\vspace{1mm}
\noindent
\textbf{Underwater Cyber-physical Security.}
Maritime cyber-physical security generally focuses on defending communication systems~\cite{davenport2015submarine, beckman2014protecting} and underwater acoustic networks~\cite{zuba2015vulnerabilities, zuba2011launching, dong2014study, xiao2015experimental}.
Researchers studying physical attacks in the underwater domain generally focus on directly tampering with components (e.g., damaging communication cables~\cite{beckman2014protecting}) or spoofing using in-band signal injection (e.g., spoofing acoustic communication between underwater network devices by generating a malicious acoustic signal~\cite{zuba2011launching}). While a recent position paper evaluated acoustic injection on HDDs~\cite{sheldon2023deep} in underwater scenarios, the research was limited to cause DoS on a single consumer-grade HDD and did not include an in-depth analysis of the attacker's capabilities or suggest defensive measures.
Our acoustic injection attack, on the other hand, evaluates complex fault-tolerant RAID systems consisting of enterprise HDDs and fine-grained control over data center resource allocation and database and distributed filesystem performance. We also evaluate the efficacy of commonly suggested defenses and propose a novel defense against underwater acoustic injection attacks.
\section{Conclusion}
We evaluate an attacker's ability to use acoustic injection to gain fine-grained control over the performance of fault-tolerant storage systems used in data centers. We deploy a submerged enclosure and test our attack in laboratory and open-water scenarios. Through this evaluation, we identify threats to distributed systems and data center management tools such as distributed filesystems, databases, and data center resource managers. We also describe limitations in commonly suggested defenses against acoustic injection and suggest a novel proof-of-concept ML-based defense which reaches 0\% False Positive Rate and 98.2\% True Positive Rate in our testbed scenario. Furthermore, we provide a theoretical analysis of the sound-induced vibrations and simulate the effect of our attack in realistic UDC deployment scenarios. We hope that our attack characterization and suggested defense improve the security of underwater data centers, which have recently emerged as a novel environmentally sustainable cloud computing technology. 

\section*{Acknowledgements}
 
We thank the anonymous shepherd and reviewers for their valuable comments. This research was supported in part by the National Science Foundation (NSF) under CNS-2055014, the Air Force Office for Scientific Research under FA8650-19-1-1741 and FA8650-19-1-0169, gifts from Meta and Texas Instruments, and JST CREST JPMJCR23M4. Any opinions, findings, and conclusions or recommendations expressed in this material are those of the authors and do not necessarily reflect the views of the NSF.

\bibliographystyle{IEEEtran}
\bibliography{ref}

\end{document}